\documentclass[12pt,a4paper]{article}

\usepackage{amsmath}
\usepackage{amssymb}
\usepackage{graphicx}
\usepackage{cite}
\usepackage{hyperref}

\begin{document}
\title{Thin and thick bubble walls II: expansion in the wall width}
\author{
	Ariel M\'{e}gevand\thanks{Member of CONICET, Argentina. E-mail address:
		megevand@mdp.edu.ar}~ 
	and Federico Agust\'{\i}n Membiela\thanks{Member of CONICET, Argentina. E-mail
		address: membiela@mdp.edu.ar} \\[0.5cm]
	\normalsize \it IFIMAR (CONICET-UNMdP)\\
	\normalsize \it Departamento de F\'{\i}sica, Facultad de Ciencias Exactas
	y Naturales, \\
	\normalsize \it UNMdP, De\'{a}n Funes 3350, (7600) Mar del Plata, Argentina }
\date{}
\maketitle

\begin{abstract}
We study the dynamics of a cosmological bubble wall beyond the approximation of an infinitely thin wall.
In a previous paper, we discussed the range of validity of this approximation
and estimated the first-order corrections due to the finite width.
Here, we introduce a systematic method to obtain the wall equation of motion and its profile at each order in the wall width.
We discuss in detail the next-to-next-to-leading-order terms. 
We use the results to treat the growth of spherical bubbles and the evolution of small deformations of planar walls.
\end{abstract}

\section{Introduction}

The motion of bubble walls in a first-order cosmological phase transition has several
consequences, such as the formation of gravitational
waves \cite{h86,tw90,ktw92,kkt94} 
or the generation of the baryon asymmetry of the universe
\cite{krs85,ckn93}. 
Generally, these interfaces
can be described by a field configuration
in which a scalar field $\phi$ varies between two values
corresponding
to two minima of the effective potential $V(\phi)$. 
Such a configuration
is similar to a domain wall \cite{vs00}. However, in the case of a domain wall,
the minima of $V$ correspond to two true vacua, while in a phase
transition the wall separates the true vacuum, which we will denote $\phi_{-}$,
from a false vacuum $\phi_{+}$. Therefore, for a domain wall,
the potential takes the same value on both sides of the interface, while
for a bubble wall we have two different values $V_{-}<V_{+}$. 
This pressure difference between phases is relevant to the dynamics
and also affects the effective treatment of the wall as a thin interface. 

Either for a bubble wall or a domain wall, it is common to use the thin-wall
approximation to derive an effective equation of motion (EOM) 
for the wall and solve separately the equation for the wall profile.
For that aim, it is usual to assume that the field only varies in
the direction perpendicular to the wall hypersurface. 
Therefore, $\phi$ is a function of a single variable $n$.
The function $\phi(n)$ describes the kink profile.
On the other hand, the surface representing the wall position 
can be parametrized as $x^{\mu}=X^{\mu}(\xi^{a})$, where $\xi^a$, $a=0,1,2$,
are two space parameters and a time parameter. 
The function
$X^{\mu}(\xi^{a})$ describes the evolution of the wall as a surface.

In the case of a domain wall, the general equation for $X^{\mu}(\xi^{a})$
has been widely discussed, even beyond the thin-wall approximation \cite{w89thick,w89collapse,ghg90,gg90,g91,sm93,l93,cg95,al94,a95,a95b,a98}.
On the other hand, for a bubble wall, only planar, cylindrical, and
spherical walls have been considered (see, e.g., \cite{lm11}), as well as small perturbations
of these cases \cite{gv91,afw90}, and generally in the limit of an infinitely thin wall.
A recently formed bubble is spherical, and, intuitively, 
the thin-wall approximation will be valid when the bubble grows and its radius
$R$ becomes much larger than the wall width $l$.
However, a bubble wall may acquire a small radius of curvature locally due to the unstable growth of small perturbations.
These can arise, for instance, from hydrodynamic instabilities 
\cite{l92,hkllm93,mm14,mm14b} or collision with other bubbles \cite{bbm1,bbm2,bbm3}.
More importantly, what matters is actually the curvature of the worldvolume \cite{w89thick,ghg90}.
Even for  a spatially planar wall,
this hypersurface will curve when accelerated by a potential difference $\Delta V=V_{+}-V_{-}$ \cite{mm23}.
We shall generally denote $L$ the length scale associated to the local radius of curvature.
We remark that the quantities $l$ and $L$ are compared in a coordinate system associated to the wall.
In particular, these lengths are not Lorentz-contracted. Indeed, 
in the usual thin-wall approximation,
the profile $\phi(n)$ is a fixed function, and its width $l$ is a constant.

The methods used for domain walls can be adapted
to bubble walls. 
In our previous work \cite{mm23}, we derived the EOM for a bubble wall of arbitrary shape 
and discussed the  finite-width corrections.
The potential difference between the minima, $\Delta V$, introduces subtleties that make the
treatment more complex. For example, the finite-width corrections for a domain wall are
of order $(l/L)^{2}$, while for a bubble wall, there are corrections of order $l/L$.
In the present paper, we will develop a perturbative
method to obtain both the EOM and the profile at any order,
and we will discuss in detail the corrections up to order $(l/L)^{2}$. 
We will consider for simplicity the case of 
a so-called vacuum transition, where the presence
of the plasma is negligible. The method is generalizable, and in a separate
paper we will incorporate the fluid into the discussion.

The plan of the paper is the following. In Sec.~\ref{apbasica}, we
briefly review the basic thin-wall approximation and its shortcomings.
We also introduce some tools that we will use in our analysis, such as
normal Gaussian coordinates and the extrinsic curvature tensor of
a hypersurface. Appendix \ref{sec:gaussian} contains additional details.
In Sec.~\ref{sec:pert_method}, we develop our method to 
obtain the wall profile and EOM	
at any order in the wall width.
In Sec.~\ref{sec:EOM}, we consider the wall motion for a spherical bubble
and a planar wall with small deformations, and
in Sec.~\ref{sec:ejemplos}
we apply the results to a specific potential. 
Analytic expressions for this potential as well as more general
expressions can be found in App.~\ref{sec:perfil}.
In Sec.~\ref{sec:nucleation},
we consider a convenient modification of our method to treat the O(3,1) invariant solution or, equivalently,
the O(4) invariant instanton.
Finally, in Sec.~\ref{sec:conclu}, we summarize our conclusions.

\section{The thin-wall approximation}

\label{apbasica}

Since the wall is not infinitely thin, we must consider a reference
surface $S$ representing its position.
For example, the set of points where $\phi$
takes a given value between $\phi_{-}$ and $\phi_{+}$. 
We will consider a  convenient definition of this locus later.
The spacetime history of the surface $S$ constitutes a hypersurface  $\Sigma$ that
can be represented explicitly by a parametrization $x^{\mu}=X^{\mu}(\xi^{a})$.
A gauge fixing is necessary to fully determine the function $X^{\mu}(\xi^{a})$.
When needed, we will use the Monge parametrization 
$x^{3}=x_{w}^{3}(x^{0},x^{1},x^{2})$, which is usually appropriate for a bubble wall \cite{mm23}.
In this case, we have $X^{\mu}(\xi^{a})=(\xi^{a},x_{w}^{3}(\xi^{a}))$.
We will also consider an implicit definition of the hypersurface, $F(x^{\mu})=0$.
For the Monge parametrization, we have $F=x^{3}-x_{w}^{3}(x^{a})$.
The normal vector to $\Sigma$,
which satisfies the conditions $N_{\mu}\partial_{a}X^{\mu}=0$ and $N_{\mu}N^{\mu}=-1$, can be defined by
\begin{equation}
N_{\mu}(x^{\nu}) = -\partial_{\mu}F/s\,\quad\textrm{with}\quad s=\sqrt{|F_{,\mu}F^{,\mu}|}.
\label{eq:Nmu}
\end{equation}
In the Monge gauge, we have 
$
N_{\mu}=(\partial_{a}x_{w}^{3},-1)/s
$
and $s^{2}=-g^{33}+2g^{3a}\partial_{a}x_{w}^{3}-g^{ab}\partial_{a}x_{w}^{3}\partial_{b}x_{w}^{3}$.

For a vacuum transition, the equation of motion for the scalar field is 
\begin{equation}
\nabla_{\mu}\nabla^{\mu}\phi+V'(\phi)=0. \label{eq:eccampo}
\end{equation}
The basic idea of the thin-wall approximation is to obtain, from Eq.~(\ref{eq:eccampo}),
an equation for the hypersurface  $X^{\mu}(\xi^{a})$,
as well as an equation for the field profile $\phi(n)$,
where $n$ is a coordinate perpendicular to $\Sigma$. 
Since the parameters $\xi^a$, $a=0,1,2$, represent coordinates on the worldvolume $\Sigma$,
it is convenient to use the coordinates $\bar{x}^{\mu}=(\xi^{a},n)$. 
In these coordinates, the field equation becomes
\begin{equation}
\partial_{n}^{2}\phi-K\partial_{n}\phi-D_{a}D^{a}\phi=V'(\phi),\label{ecfi}
\end{equation}
where we have defined the quantity $K=-\bar{g}^{ab}\bar{\Gamma}_{ab}^{n}$
and used the notation 
\begin{equation}
D_{a}D^{a}\phi\equiv\bar{g}^{ab}\left(\partial_{a}\partial_{b}\phi-\bar{\Gamma}_{ab}^{c}\partial_{c}\phi\right).
\end{equation}

There is not a unique way to assign coordinates $\xi^{a},n$ to points away from $\Sigma$,
and we shall specifically use normal Gaussian coordinates, which are
constructed as follows. We consider the set of geodesics that cross
$\Sigma$ perpendicularly. 
In a certain neighborhood of $\Sigma$, any point $x^{\mu}$ lies on one of these geodesics. 
We define the coordinate $n$ as the distance to $\Sigma$ along the geodesic.
The point $x^\mu$ can be determined by $n$ and the point $X^{\mu}(\xi^{a})$
where the geodesic intersects $\Sigma$.
We can thus assign to $x^\mu$ the new coordinates $\bar{x}^{\mu}=(\xi^{a},n)$. 
Near $\Sigma$, the change of coordinates is given by the expansion
\begin{equation}
x^{\mu}=X^{\mu}(\xi^{a})+N^{\mu}(\xi^{a})n+\cdots,\label{eq:Gaussian}
\end{equation}
If we stay very close to $\Sigma$, this change of coordinates is
equivalent to the simpler one where the expansion (\ref{eq:Gaussian})
is truncated at the linear order in $n$. However, normal Gaussian coordinates
have good properties that are particularly useful if we go beyond
that order, which is necessary when dealing with a thick wall. 
A few of these properties, which we will use in the next sections, are the following
(see App.~\ref{sec:gaussian} for details and further expressions).
The metric tensor in the coordinates $\bar{x}^{\mu}=(\xi^{a},n)$ fulfills 
$\bar{g}_{nn}=-1 $, $ \bar{g}_{an}=0.$
In particular, this holds for the induced metric on $\Sigma$, $\gamma^{ab}=\bar{g}^{ab}|_{n=0}$.
The Christoffel symbols in normal Gaussian coordinates are given by
\begin{equation}
\bar{\Gamma}_{nn}^{\mu}=\bar{\Gamma}_{\mu n}^{n}=0,
\quad
\bar{\Gamma}_{ab}^{n}=\frac{1}{2}\partial_{n}\bar{g}_{ab},
\quad
\Gamma_{na}^{b}=\frac{1}{2}g^{bc}\partial_{n}\bar{g}_{ca}.
\label{eq:GammaGauss}
\end{equation}
We can invert the coordinate transformation (\ref{eq:Gaussian}) order
by order to return to the original coordinates. We are only interested in the expression for $n(x^{\mu})$,
\begin{equation}
n=\frac{F}{s} + \frac{N^{\nu}\partial_{\nu}s}{2s} \left(\frac{F}{s}\right)^{2} + \left[\frac{\left(N^{\nu}\partial_{\nu}s\right)^{2}}{2s^{2}}
+\frac{N^{\rho}N^{\nu}N^{\mu}\nabla_{\rho}\nabla_{\nu}(sN_{\mu})}{6s}\right] \left(\frac{F}{s}\right)^{3}+\cdots.
\label{eq:n_orden2}
\end{equation}

The quantity $K$ appearing in the field equation, Eq.~(\ref{ecfi}), is
related to the extrinsic curvature of the hypersurfaces
of $n=$ constant, which we denote $\Sigma_{n}$. 
Indeed, the extrinsic curvature tensor can be defined
as $K_{\mu\nu}=-\nabla_{\mu}n_{\nu}$, where $n^{\mu}$ is the normal
vector to $\Sigma_{n}$, which in normal Gaussian coordinates takes
the form $\bar{n}^{\mu}=(0,0,0,1)$ (see App.~\ref{sec:gaussian}
for more details). Hence, we have 
$\bar{K}_{n\mu}=0$, $\bar{K}_{ab}=-\bar{\Gamma}_{ab}^{n}$,
and 
\begin{equation}
K=\bar{g}^{ab}\bar{K}_{ab}=g^{\mu\nu}K_{\mu\nu}.\label{eq:K}
\end{equation}
The quantity $K$ is sometimes called the mean curvature. In particular,
at $n=0$ we have $\Sigma_{n}=\Sigma$ and $n^\mu = N^{\mu}$. 
In terms of our vector field $N_{\mu}$, the extrinsic curvature tensor of $\Sigma$ is given by
\begin{equation}
K_{\mu\nu}=-{h_{\mu}}^{\rho}\nabla_{\rho}N_{\nu}.
\label{Kmunu}
\end{equation}
where ${h_{\mu}}^{\nu}$ is the projection tensor orthogonal to $N_{\mu}$
and tangent to $\Sigma$ \cite{carroll}, 
\begin{equation}
{h_{\mu}}^{\nu}\equiv\left(\delta_{\mu}^{\nu}+N_{\mu}N^{\nu}\right).
\label{eq:proyector}
\end{equation}
For our perturbative treatment, we are going to need the derivatives
of $K$ evaluated at $n=0$. We shall use the notation $K|_{n=0}$
or just $K|_{0}$ for $K(\xi^{a},0)$. 
Applying the operator $\partial_{n}=n^{\mu}\nabla_{\mu}$ to the previous expressions,
we obtain (see App.~\ref{sec:gaussian})
\begin{equation}
K|_{n=0}=-N_{;\mu}^{\mu},\quad\partial_{n}K|_{n=0}=N_{;\nu}^{\mu}N_{;\mu}^{\nu}\quad\partial_{n}^{2}K|_{n=0}=-2N_{;\nu}^{\mu}N_{;\rho}^{\nu}N_{;\mu}^{\rho}.\label{eq:Kcovar}
\end{equation}

The so-called thin-wall approximation actually involves a series of approximations.
The first step is to assume that the field $\phi(\xi^{a},n)$ only
depends on the variable $n$ (see, e.g., \cite{vs00}). 
This assumption eliminates the derivatives
tangent to the hypersurface in Eq.~(\ref{ecfi}), so we have
\begin{equation}
\partial_{n}^{2}\phi-K\partial_{n}\phi=V'(\phi).
\label{eq:primerasup}
\end{equation}
By consistency, this equation also requires $K$ to depend only
on $n$ \cite{sm93}, $K(\xi^{a},n)=K(n)$. Evaluating this condition
at $n=0$ and using the expression for $K$ from Eq.~(\ref{eq:Kcovar}), we obtain an equation
for the surface, namely, $N_{;\mu}^{\mu}=$ constant. This
is indeed the form of the effective EOM for the wall in the basic
thin-wall approximation. 
However, taking the derivative of this condition,
evaluating again at $n=0$, and using the expression for $\partial_n K$ from
Eq.~(\ref{eq:Kcovar}), we obtain a different EOM. This shows that (except for particular
cases) the assumption $\phi=\phi(n)$ must cease to hold as soon as
we move away from the reference hypersurface $\Sigma$. In Ref.~\cite{mm23},
we called the approximation $\partial_{a}\phi=0$ the ``incompressibility assumption'' and
argued that the error made by neglecting $\partial_{a}\phi$ versus $\partial_{n}\phi$ is 
\begin{equation}
\frac{\partial_{a}\phi}{\partial_{n}\phi}\sim\frac{\delta l}{L}<\frac{l}{L},
\end{equation}
where $\delta l$ represents the variations in the wall width $l$,
while $L$ is the local curvature radius of $\Sigma$. 
In \cite{mm23}, we argued that $\delta l\ll l$ is to be expected. 
Even in the case $\delta l\sim l$, the terms with derivatives with
respect to $\xi^{a}$ in Eq.~(\ref{ecfi}) are of order $(l/L)^{2}$
with respect to the leading term $\partial_{n}^{2}\phi$. 
We shall see that we actually have  $\delta l\sim l^{2}$ and the term $D_{a}D^{a}\phi$
is of order $(l/L)^{4}$ with respect to $\partial_{n}^{2}\phi$.

The next assumption that is usually made in the thin-wall
approximation is that $K(n)$ varies little over the width of the
wall,  where $\partial_{n}\phi\neq 0$.
Thus, multiplying Eq.~(\ref{eq:primerasup})
by $\partial_{n}\phi$ and integrating from $n=-\infty$ to $n=+\infty$,
we obtain 
\begin{equation}
K=-\Delta V/\sigma,\label{eq:EOMbasica}
\end{equation}
where $\Delta V=V_{+}-V_{-}$ with $V_\pm =V(\phi_\pm)$, and
$\sigma=\int_{-\infty}^{+\infty}(\partial_{n}\phi)^{2}dn$ is the surface tension.
Using Eq.~(\ref{eq:Kcovar}), we obtain an equation for the normal
vector, ${N^{\mu}}_{;\mu}=\Delta V/\sigma$. Using Eq.~(\ref{eq:Nmu})
with the Monge parametrization, we obtain an equation for the wall position
$x_{w}^{3}(x^{0},x^{1},x^{2})$. This equation depends on the chosen
coordinates and we will discuss it in Sec.~\ref{sec:EOM}.
From a geometric point of view, the meaning of Eq.~(\ref{eq:EOMbasica}) is that
the potential difference causes an acceleration and, therefore, curves the hypersurface.

To obtain the value of $\sigma$, we need to solve the equation for
$\phi$, Eq.~(\ref{eq:primerasup}), which depends on $K$ and, hence, on $\sigma$. Therefore,
it is usual to make yet another approximation, which consists in neglecting
the second term in Eq.~(\ref{eq:primerasup}) since it is of order
$l/L$ with respect to the first one. The resulting equation, $\partial_{n}^{2}\phi=V'(\phi)$,
is readily integrated. Imposing the condition $\phi(+\infty)=\phi_{+}$,
we obtain 
\begin{equation}
\phi^{\prime2}=2\left(V-V_{+}\right).\label{eq:ecfibasica}
\end{equation}
If we evaluate this equation at $n=-\infty$, we obtain $\Delta V=0$.
This means that this approximation requires, for consistency, a potential
with $V_{+}=V_{-}$.
This looks incompatible with Eq.~(\ref{eq:EOMbasica}), but here we have made an extra 
approximation to estimate the parameter $\sigma$.
In any case,
as discussed in \cite{mm23}, 
the approximation of small $l/L$ may break down since
Eq.~(\ref{eq:EOMbasica}) introduces a curvature radius $L$ which
is naturally of order $l$
unless $\Delta V$ is relatively small. 
Specifically, for a planar wall we roughly have $l/L \sim \Delta V/V_{b}$,
where $V_b$ is the height of the barrier between the minima
of the potential%
\footnote{This estimation is valid for a nearly degenerate potential \cite{mm23}, where the barrier height can be unambiguously defined as the difference between the maximum of the potential and any of the minima, and we have $\Delta V\ll V_b$.
	In Sec.~\ref{sec:ejemplos}, we will use the ratio $\Delta V/V_b$ to characterize the shape of the potential even when this condition is not fulfilled. 
	We will use the definition $V_b=V(\phi_b)-V_+$, where $\phi_b$ is the value of $\phi$ at the maximum.}
(additionally, the wall may be curved even for $\Delta V=0$).

To implement the approximation of a small $\Delta V$, it is convenient
to write $V$ in the form \cite{c77}
\begin{equation}
V(\phi)=V_{0}(\phi)+\tilde{V}(\phi),\label{eq:potpert}
\end{equation}
where $V_{0}$ is a degenerate potential and $\tilde{V}$ causes the
energy difference between minima. The decomposition (\ref{eq:potpert})
is not unique and a convenient form for
$\tilde{V}$ can be chosen  (e.g., a linear term $\tilde{V}=c\phi$
or a quadratic term $\tilde{V}=\frac{1}{2}\delta m^{2}\phi^{2}$).
Then, we define $V_{0}=V-\tilde{V}$ and we choose the
parameters in $\tilde{V}$ so that the minima of $V_{0}$
have the same energy. Denoting these minima $a_{\pm}$,
we have the conditions \cite{mm23}
\begin{align}
\tilde{V}'(a_{\pm})&=V'(a_{\pm}),\label{eq:condminV0}
\\
\tilde{V}(a_{+})-\tilde{V}(a_{-})&=V(a_{+})-V(a_{-}).\label{eq:condV0deg}
\end{align}
If $\tilde{V}$ contains a single free parameter, these equations
determine its value, which will be of the order of $\Delta V$.
Using the approximation $V\simeq V_{0}$, Eq.~(\ref{eq:ecfibasica})
can be easily solved, and from the profile $\phi(n)$ we obtain the
surface tension $\sigma$. Using this value in Eq.~(\ref{eq:EOMbasica}),
we obtain a closed equation for the wall surface. 
It is worth remarking
that this equation of motion for the wall and the equation for the
profile, Eq.~(\ref{eq:ecfibasica}), were both derived from Eq.~(\ref{eq:primerasup}).
However, in Eq.~(\ref{eq:EOMbasica}), the leading-order term $\partial_{n}^{2}\phi$
vanished upon integration, so the whole equation (\ref{eq:EOMbasica})
is of higher order. In contrast, Eq.~(\ref{eq:ecfibasica}) retains
this term and neglects the second one in Eq.~(\ref{eq:primerasup}),
which, according to Eq.~(\ref{eq:EOMbasica}), is proportional to
$\Delta V$. This is why Eq.~(\ref{eq:ecfibasica})  is consistent with $\Delta V=0$.

\section{The perturbative expansion}

\label{sec:pert_method}

To go beyond the above approximations, let us consider the first integral of Eq.~(\ref{ecfi}), 
\begin{equation}
\frac{1}{2}(\partial_{n}\phi)^{2} + 
\int_{n}^{\infty}\left[K(\partial_{n}\phi)^{2}+\partial_{n}\phi D_{a}D^{a}\phi\right]d\tilde n
=-\left[V_{+}-V(\phi)\right], \label{eq:primeraint}
\end{equation}
which is obtained by multiplying the equation by $\partial_{n}\phi$ and integrating through the wall.
We have imposed the condition $\phi(+\infty)=\phi_{+}$.
We also assume that $\phi(-\infty)=\phi_{-}$, so, evaluating Eq.~(\ref{eq:primeraint}) at $n=-\infty$, we have 
\begin{equation}
\int_{-\infty}^{+\infty}\left[(\partial_{n}\phi)^{2}K+\partial_{n}\phi D_{a}D^{a}\phi\right]dn=-\Delta V.\label{eq:eom}
\end{equation}
As we will see below, these equations are the generalization of Eqs.~(\ref{eq:EOMbasica}) and (\ref{eq:ecfibasica})
to all order in the wall width.
We have not yet given a precise definition of the reference surface $S$ associated to the wall position, 
whose worldvolume is given by the hypersurface $\Sigma$ located at $n=0$.
Taking into account that the energy density of the wall is proportional to the quantity $(\partial_n\phi)^2$,
it makes sense to define the wall position as the average of $n$ weighted with this quantity.
Since, by definition, the wall position is $n=0$, we have
\begin{equation}
\int_{-\infty}^{+\infty}(\partial_{n}\phi)^{2}\,n\,dn=0 . \label{eq:cero_n}
\end{equation}
This condition will fix an integration constant in the solution of Eq.~(\ref{eq:primeraint}).
We can also define the wall width $l$ as the root mean square value of $n$, i.e.,
\begin{equation}
	l^2 = \frac{\int_{-\infty}^{+\infty}(\partial_{n}\phi)^{2}n^{2}dn}{\int_{-\infty}^{+\infty}(\partial_{n}\phi)^{2}dn}
	\equiv \frac{\mu}{\sigma},
	\label{eq:defl}
\end{equation}
where $\sigma=\int_{-\infty}^{+\infty}(\partial_{n}\phi)^{2}dn$ is the surface tension and we have also defined
the quantity $\mu=\int_{-\infty}^{+\infty}(\partial_{n}\phi)^{2}n^{2}dn$, which will appear below.

Let us denote $\phi_{0}$ the solution obtained
from Eq.~(\ref{ecfi}) or its first integral (\ref{eq:primeraint}) with the approximations 
used in the previous section, namely, $D_{a}D^{a}\phi=0$,
$K=0$, and $V=V_{0}$. Thus, Eq.~(\ref{ecfi}) becomes
\begin{equation}
	\phi_{0}''(n)=V_{0}'(\phi_{0})\label{eq:ecfi0}
\end{equation}
and Eq.~(\ref{eq:primeraint}) becomes $\phi_0^{\prime2}=2(V_0-V_{0+})$, 
which is the usual thin-wall equation for the profile, Eq.~(\ref{eq:ecfibasica}),
with $V$ replaced by $V_0$ and $V_+$ by $V_{0+}\equiv V_0(a_+)$, i.e.,
for consistency, 
$\phi_{0}$ must fulfill the boundary conditions $\phi_{0}(\pm\infty)=a_{\pm}$
(the minima of $V_0$).
We regard this equation as the leading-order version of Eq.~(\ref{eq:primeraint}).
We thus have
\begin{equation}
	\phi_{0}^{\prime}(n)=-\sqrt{2\left[V_{0}(\phi_{0})-V_{0}(a_{+})\right]}\equiv -\phi_h(\phi_0),
	\label{eq:primeraint0}
\end{equation}
where the sign chosen for the square root corresponds to assuming $a_{+}<a_{-}$.
Integrating Eq.~(\ref{eq:primeraint0}),
we obtain the implicit solution
\begin{equation}
n=-\int_{\phi_{*}}^{\phi_{0}}\frac{d\phi}{\phi_h(\phi)}+n_{*}.\label{eq:fi0}
\end{equation}
The  value $\phi_*$ is arbitrary, and we determine the constant $n_{*}$ by imposing the condition (\ref{eq:cero_n}) to $\phi_0$.
We obtain
\begin{equation}
n_{*}=\sigma_{0}^{-1}\int_{a_{+}}^{a_{-}}d\phi_{0}\left[\phi_{h}(\phi_{0})\int_{\phi_{*}}^{\phi_{0}}\frac{d\phi}{\phi_{h}(\phi)}\right],
\label{eq:nast}
\end{equation}
with $\sigma_0=\int_{-\infty}^{+\infty}(\partial_{n}\phi_0)^{2}dn$.
Notice that using the same approximations in Eq.~(\ref{eq:eom}) just gives
$0=0$. Nevertheless, at the next order we will obtain the leading-order equation for the wall, Eq.~(\ref{eq:EOMbasica}).

To obtain the field profile and the wall EOM at higher orders in the wall width,
we need to take into account that 
the quantities $K$, $D_{a}D^{a}\phi$, and $V-V_0$ do not vanish.
We begin by writing the field in the form
\begin{equation}
\phi=\phi_{0}+\phi_{1}+\phi_{2}+\cdots,\label{eq:expanfi}
\end{equation}
where each term is of order $l/L$ higher than the previous one. The
concrete definition of each term $\phi_{i}$ is that the sum (\ref{eq:expanfi})
up to each order i gives the solution to Eq.~(\ref{eq:primeraint})
to that order. In particular, $\phi_{0}$ is the solution of Eq.~(\ref{eq:primeraint0}).
These solutions will include the boundary conditions at each order,
$\phi(\pm\infty)=\phi_{\pm}$, so we write
\begin{equation}
\phi_{\pm}=a_{\pm}+\phi_{1\pm}+\phi_{2\pm}+\cdots.
\end{equation}
Taking into account that the term $\tilde{V}$ in Eq.~(\ref{eq:potpert})
is of order $l/L$ and expanding $V$ in powers of $\phi-\phi_{0}$,
we have
\begin{equation}
V(\phi)=V_{0}(\phi_{0}) + \big[\tilde{V}(\phi_{0})+V_{0}'(\phi_{0})\phi_{1} \big]+
\big[\tilde{V}'(\phi_{0})\phi_{1}+{\textstyle\frac{1}{2}}V_{0}''(\phi_{0})\phi_{1}^{2}+V_{0}'(\phi_{0})\phi_{2}\big]+\cdots,
\end{equation}
where we have grouped terms of the same order in $l/L$. In particular,
evaluating this expression at the minima, we obtain
$V_\pm = V_{0\pm}+V_{1\pm}+\cdots$, with
\begin{align}
	V_{0\pm} & =  V_0(a_\pm),  \qquad
	V_{1\pm}= \tilde{V}(a_{\pm}), \qquad
	V_{2\pm}=  \tilde{V}'(a_{\pm})\phi_{1\pm} + {\textstyle\frac{1}{2}} V_{0}''(a_{\pm})\phi_{1\pm}^{2}   \label{V1}
	\\
	V_{3\pm} &= {\textstyle\frac{1}{2}} \tilde{V}''(a_{\pm})\phi_{1\pm}^{2} + {\textstyle\frac{1}{6}} V_{0}'''(a_{\pm})\phi_{1\pm}^{3} + V_{0}''(a_{\pm})\phi_{1\pm}\phi_{2\pm} +\tilde{V}'(a_{\pm})\phi_{2\pm} , \, \ldots \label{V3} 
\end{align}
and  $\Delta V=\Delta V_{1}+\cdots$, with
$\Delta V_i =V_{i+}-V_{i-}$. 

In Sec.~\ref{apbasica}, we considered two different approximations for the quantity $K$, namely, 
that it can be neglected, and that its variation can be neglected.
We must regard these as approximations of different order in the wall width.
First of all, let us write
\begin{equation}
K=K_{0}+K_{1}+K_{2}+\cdots,
\label{expanK}
\end{equation}
where each term is of order $l/L$ higher than the previous one.
Unlike $\phi$, $K$ does not have a significant variation within
the width of the wall, so we can use its expansion as a Taylor series in $n$. 
We consider such an expansion for each term in Eq.~(\ref{expanK}),
\begin{equation}
K_{i}=K_{i}|_{n=0} + \partial_{n}K_{i}|_{n=0}\,n + \frac{1}{2}\partial_{n}^{2}K_{i}|_{n=0}\,n^{2} +\cdots ,
\label{eq:expanKi}
\end{equation}

It is important to note that the subscripts in the expansions (\ref{eq:expanfi})-(\ref{eq:expanKi}) only
indicate the relative order of a term within the quantity being expanded.
We need to insert these expansions in Eqs.~(\ref{eq:primeraint})-(\ref{eq:eom})
and compare the terms. Each of the above quantities scales with different
powers of $l$ and $L$, and, besides, there are derivatives and integrals in Eqs.~(\ref{eq:primeraint})-(\ref{eq:eom}). 
To assess the absolute order of
a term, we must take into account, in the first place, that the order of magnitude of both
$\phi$ and $l$ is given  by the scale of the theory $v$ as $\phi\sim v$, $l\sim v^{-1}$.
This means that we have $\phi_{0}\sim l^{-1}$,
$\phi_{1}\sim l^{-1}(l/L)=L^{-1}$, $\phi_{2}\sim lL^{-2}$, and so on.
In the second place, the mean curvature $K$ is (by definition of
$L$) of order $L^{-1}$, so we have
$K_{0}\sim L^{-1}$, $K_{1}\sim L^{-1}(l/L)=lL^{-2}$, and so
on. In the third place, a derivative $\partial_{n}$ applied to $\phi_{i}$
changes the order by a factor $l^{-1}$, but applied to $K$ changes
the order by $L^{-1}$. On the other hand, the operator $D_{a}D^{a}$
is of order $L^{-2}$. Finally, we must take into account that
the integral increases the order by a factor $l$, since the
integrand vanishes outside the wall. For the same reason,
each power of $n$ increases the order by a factor $l$. 
The lowest-order terms in Eq.~(\ref{eq:primeraint}) are of order
$l^{-4}$. Retaining only these terms,
we obtain Eq.~(\ref{eq:primeraint0}).
The terms of order $i$ are those of order $l^{-4}(l/L)^i$.

When we insert the expansion (\ref{expanK})-(\ref{eq:expanKi}) in
Eqs.~(\ref{eq:primeraint}) and (\ref{eq:eom}), the integral of $K(\partial_{n}\phi)^{2}$ will split into a combination of integrals
of the form
\begin{equation}
I^{(k)}(\xi^{a},n)=\int_{n}^{\infty}(\partial_{n}\phi)^{2}\tilde{n}^{k}\,d\tilde{n},
\quad
\bar{I}^{(k)}(\xi^{a})=\int_{-\infty}^{+\infty}(\partial_{n}\phi)^{2}{n}^{k}\,d{n}
.\label{eq:GHI}
\end{equation}
Thus, we have $I^{(k)}|_{n=+\infty}=0$ and $I^{(k)}|_{n=-\infty}=\bar{I}^{(k)}$. 
In particular, for $k=0,1,2$, we obtain the integrals appearing in Eqs.~(\ref{eq:cero_n})-(\ref{eq:defl}),
so we have
$\sigma=\bar{I}^{(0)}$, $\mu =\bar{I}^{(2)}$, and the condition $\bar{I}^{(1)}=0$.%
\footnote{Since $\bar{I}^{(1)}=0$ and $\partial_{n}I^{(1)}=-n\left(\partial_{n}\phi\right)^{2}$,
	an integration by parts gives the relation $\int_{-\infty}^{+\infty}I^{(1)}n^jdn=\frac{1}{j+1}\bar{I}^{(j+2)}$,
	which can be used to obtain some of the results we give below.}
When we insert the expansion (\ref{eq:expanfi}) in the quantities
(\ref{eq:GHI}), we will have, at each order in $l/L$, integrals
of the form
\begin{equation}
I_{0}^{(k)}=\!\int_{n}^{\infty}\!\!\! (\partial_{n}\phi_{0})^{2}\tilde{n}^{k}d\tilde{n},\,
I_{1}^{(k)}=\!\int_{n}^{\infty}\!\!\! 2 \partial_{n}\phi_{0}\partial_{n}\phi_{1}\tilde{n}^{k}d\tilde{n},\,
I_{2}^{(k)}=\!\int_{n}^{\infty}\!\!\! [(\partial_{n}\phi_{1})^{2}+2\partial_{n}\phi_{0}\partial_{n}\phi_{2}]\tilde{n}^{k}d\tilde{n},
\label{eq:Gi}
\end{equation}
and so on, as well as definite integrals $\bar{I}^{(k)}_i$.
In particular, we have
$\bar{I}_{i}^{(0)}|=\sigma_{i}$ and $\bar{I}_{i}^{(2)} = \mu_{i}$, where $\sigma_{i}$ and $\mu_{i}$ are
the terms of the expansion of $\sigma$ and $\mu$ at successive orders
in $l/L$.

Expanding Eq.~(\ref{eq:primeraint}) to order $i$ and taking into account that the field $\phi_{i-1}$ fulfills
the equation for the previous order, we obtain the equation for $\phi_i$,
\begin{equation}
\partial_{n}\phi_{0}\partial_{n}\phi_{i}-\partial_{n}^{2}\phi_{0}\phi_{i}=f_{i},
\label{eq:primeraintpert}
\end{equation}
where we have also used Eq.~(\ref{eq:ecfi0}). The first few functions $f_{i}$
are given by
\begin{align}
f_{1}=\: & \tilde{V}(\phi_{0})- V_{1+} -K_{0}|_{0}I_{0}^{(0)}, \label{f1}\\
f_{2}=\: & \tilde{V}'(\phi_{0})\phi_{1} + {\textstyle\frac{1}{2}} V_{0}''(\phi_{0})\phi_{1}^{2} 
-V_{2+} -{\textstyle\frac{1}{2}} (\partial_{n}\phi_{1})^{2}  -K_{0}|_{0}I_{1}^{(0)}-K_{1}|_{0}I_{0}^{(0)}-\partial_{n}K_{0}|_{0}I_{0}^{(1)}, 
\label{f2}\\
f_{3} = \: & \tilde{V}'(\phi_{0})\phi_{2}  + {\textstyle\frac{1}{2}}\tilde{V}''(\phi_{0})\phi_{1}^{2} +V_{0}''(\phi_{0})\phi_{1}\phi_{2}
+ {\textstyle\frac{1}{6}}V_{0}'''(\phi_{0})\phi_{1}^{3}  -V_{3+}
 -\partial_{n}\phi_{1}\partial_{n}\phi_{2} \nonumber \\
 & -K_{0}|_{0}I_{2}^{(0)}-K_{1}|_{0}I_{1}^{(0)}-K_{2}|_{0}I_{0}^{(0)}
 -\partial_{n}K_{0}|_{0}I_{1}^{(1)}-\partial_{n}K_{1}|_{0}I_{0}^{(1)}
 -{\textstyle\frac{1}{2}}\partial_{n}^{2}K_{0}|_{0}I_{0}^{(2)}.  \label{f3}
\end{align}
Similarly, expanding Eq.~(\ref{eq:eom}), we will obtain the equations for the quantities $K_i|_0$.
Instead of repeating the procedure, we may, equivalently, evaluate Eqs.~(\ref{eq:primeraintpert})
at $n=-\infty$, which gives $f_{i}|_{n=-\infty}=0$. 
Therefore, Eqs.~(\ref{f1})-(\ref{f3}) give
\begin{align}
	K_{0}|_{n=0}\sigma_{0} & =-\Delta V_{1},\label{eom1}\\
	K_{0}|_{n=0}\sigma_{1}+K_{1}|_{n=0}\sigma_{0} & =-\Delta V_{2},\label{eom2}\\
	\sigma_{2}K_{0}|_{n=0}+\sigma_{1}K_{1}|_{n=0}+\sigma_{0}K_{2}|_{n=0}+(\mu_{0}/2)(\partial_{n}^{2}K_{0})|_{n=0} & =-\Delta V_{3}.\label{eom3}
\end{align}
The quantity $K|_0$ at lowest non-trivial order is given by Eq.~(\ref{eom1}), 
and we obtain the usual equation of motion for the wall, Eq.~(\ref{eq:EOMbasica}).
We remark that the parameter $\sigma_{0}$ and, hence, the quantity $K_0|_0$, involves the field profile
$\phi_{0}$ obtained from the previous-order equation (\ref{eq:primeraint0}). 
Therefore, the function $f_1$ only depends on the previous solution $\phi_0$, 
and is a source term in the equation for $\phi_1$.
Notice also that $f_{1}$ is only a function of $n$, so Eq.~(\ref{eq:primeraintpert}) gives a solution of the form $\phi_{1}=\phi_{1}(n)$.
The term containing  $D_{a}D^{a}\phi$ in Eqs.~(\ref{eq:primeraint})-(\ref{eq:eom}) has not appeared in Eqs.(\ref{f1})-(\ref{eom3})
because it vanishes at these low orders.
Indeed, its lowest-order terms $D_{a}D^{a}\phi_{0}$, $D_{a}D^{a}\phi_{1}$, would appear in the quantities $f_{2}$ and
$f_3$,  
but these terms vanish since $\phi_{0}$ and $\phi_1$ do not depend on $\xi^{a}$. 

To solve the system of differential equations (\ref{eq:primeraintpert}),
we only need to note that the $i$-th function $f_{i}$ depends only
on the previous solutions $\phi_{0},\ldots,\phi_{i-1}$. Once these
are solved, the equation for $\phi_{i}$ is just a first order non-homogeneous
differential equation in the variable $n$. The homogeneous equation
is the same for every $i$, namely, $\phi_{0}'\partial_{n}\phi_{i}-\phi_{0}''\phi_{i}=0$,
whose general solution is proportional to the function $\phi_h$ defined in Eq.~(\ref{eq:primeraint0}).
The general solution of the non-homogeneous equation can be obtained,
e.g., by variation of constants, and is given by
\begin{equation}
\phi_{i}(\xi^{a},n)=\phi_{h}(n)\left[C_{i}(\xi^{a},n)+c_{i}(\xi^{a})\right],\label{eq:solfii}
\end{equation}
where the function $C_{i}$ is given by
\begin{equation}
C_{i}(\xi^{a},n)=-\int_{n_{*}}^{n}\frac{f_{i}(\tilde{n},\xi^{a})}{\phi_h(\tilde{n})^{2}}d\tilde{n},\label{eq:Ci}
\end{equation}
and $c_{i}$ is a constant of integration (with respect to $n$),
which must be determined by imposing the condition (\ref{eq:cero_n})
at each order, i.e., $\bar{I}_{i}^{(1)}=0$.
Integrating by parts in Eq.~(\ref{eq:Gi}) and replacing (\ref{eq:solfii}), we obtain, for $i=1,2$,
\begin{align}
c_{1} & =-2\sigma_{0}^{-1}\int_{-\infty}^{+\infty}\left[\phi_{0}^{\prime}+n\phi_{0}''\right]\phi_{0}'C_{1}dn, \label{c1}\\
c_{2} & =-\sigma_{0}^{-1}\int_{-\infty}^{+\infty}\left[2\left(n\phi_{0}''+\phi_{0}'\right)\phi_{0}'C_{2}+\phi_1^{\prime 2} n\right]dn.\label{c2}
\end{align}

While the function $f_{1}$ only depends on $n$, the function $f_{2}$
can depend on $\xi^{a}$ since the quantity $\partial_{n}K_{0}|_{0}$ in Eq.~(\ref{f2})
can depend on the point of $\Sigma$.
Therefore,
the field $\phi_{2}$ will generally depend on $\xi^{a}$. It is convenient
to single out the $\xi^{a}$-dependent term in the expressions. Hence,
we write 
\begin{equation}
f_{2}(\xi^{a},n)=\tilde{f}_{2}(n)-I_{0}^{(1)}(n)\,\partial_{n}K_{0}(\xi^{a},0)\label{eq:f2sep}
\end{equation}
and $C_{2}=C_{2a}+C_{2b}\partial_{n}K_{0}|_{n=0}$, where
\begin{equation}
C_{2a}(n)=\int_{n_{*}}^{n}\frac{\tilde{f}_{2}(\tilde{n})}{\phi_h(\tilde{n})^{2}}d\tilde{n},\quad C_{2b}(n)=-\int_{n_{*}}^{n}\frac{I_{0}^{(1)}(\tilde{n})}{\phi_h(\tilde{n})^{2}}d\tilde{n}.\label{eq:C2ab}
\end{equation}
We also define 
\begin{equation}
c_{2a}=-\sigma_{0}^{-1} \!\! \int_{-\infty}^{+\infty}\! 
[2(n\phi_{0}''+\phi_{0}')\phi_{0}'C_{2a} +\phi_{1}^{\prime 2}n]dn,
\,
c_{2b}=-\sigma_{0}^{-1} \!\! \int_{-\infty}^{+\infty} \!\!\! 2(n\phi_{0}''+\phi_{0}')\phi_{0}'C_{2b}dn,\label{eq:c2ab}
\end{equation}
so that we can write $\phi_{2}(\xi^{a},n)=\phi_{2a}(n)+\phi_{2b}(n)\partial_{n}K_{0}(\xi^{a},0)$.
The complete wall profile up to next-to-next-to-leading order is given by
\begin{equation}
	\phi(\xi^a,n)=\phi_{0}(n)+\phi_{1}(n)+\phi_{2a}(n)+\phi_{2b}(n)\partial_{n}K_{0}(\xi^a,0).\label{eq:fitot}
\end{equation}
To change from the
coordinates $\xi^{a},n$ to the original coordinates $x^{\mu}$, we
can use Eq.~(\ref{eq:n_orden2}) to obtain $n(x^{\mu})$,
while Eq.~(\ref{eq:Kcovar}) gives the
quantity $\partial_{n}K|_{0}$ as a scalar in terms of $N^{\mu}$,
and we can readily write it in any coordinates.

To solve for the quantity $K|_{n=0}$ at each order, we first obtain, from Eq.~(\ref{eom1}),
\begin{equation}
K_{0}|_{n=0}=-\Delta V_{1}/\sigma_{0}.\label{eq:K0}
\end{equation}
Using Eq.~(\ref{eq:K0})
in Eq.~(\ref{eom2}), we can solve for $K_{1}|_{0}$,
\begin{equation}
K_{1}|_{n=0}=-\Delta V_{2}/\sigma_{0}+\Delta V_{1}\sigma_{1}/\sigma_{0}^{2}.\label{eq:K1}
\end{equation}
The quantity $\sigma_{1}$ depends on the solution $\phi_{1}(n)$
and is a constant, so the first correction to $K|_{0}$ is a constant
too. Using Eq.~(\ref{eq:primeraintpert}) in the
second of Eqs.~(\ref{eq:Gi}), we obtain $\sigma_{1}=\int_{-\infty}^{+\infty}f_{1}(n)dn$.
Finally, using Eqs.~(\ref{eq:K0}) and (\ref{eq:K1}) in Eq.~(\ref{eom3}),
we solve for $K_{2}|_{0}$,
\begin{equation}
K_{2}|_{n=0}=\Delta V_{1}\sigma_{2}/\sigma_{0}^{2}-\Delta V_{1}\sigma_{1}^{2}/\sigma_{0}^{3}+\Delta V_{2}\sigma_{1}/\sigma_{0}^{2}
-\Delta V_{3}/\sigma_{0}-(\mu_0/2\sigma_0)\partial_{n}^{2}K_{0}|_{n=0}.\label{eq:preK2}
\end{equation}
This correction to $K$
will not be a constant in general, since the quantities $\sigma_{2}$
and $\partial_{n}^{2}K_{0}|_{0}$ may depend on $\xi^{a}$. Using
Eq.~(\ref{eq:primeraintpert}) in the third of Eqs.~(\ref{eq:Gi}),
the separation (\ref{eq:f2sep}) gives
\begin{equation}
\sigma_{2}=\tilde{\sigma}_{2}-\mu_{0}\partial_{n}K_{0}(\xi^{a},0),\label{eq:sigma2sep}
\end{equation}
where the quantities $\tilde{\sigma}_{2}=\int_{-\infty}^{+\infty}(\phi_{1}^{\prime2}+\tilde{f}_{2})dn$ and
$\mu_0=\int_{-\infty}^{+\infty}\phi_0^{\prime 2}n^{2}dn$ are constants.

Using Eqs.~(\ref{eq:Kcovar}), we can write Eqs.~(\ref{eq:K0})-(\ref{eq:sigma2sep})
in terms of the normal vector $N^{\mu}$. Thus, Eq.~(\ref{eq:K0})
becomes $N_{0;\mu}^{\mu}=\Delta V_{1}/\sigma_{0}$.
This is the leading-order (LO) equation for the hypersurface and
is equivalent to Eq.~(\ref{eq:EOMbasica}). If we are satisfied
with the lowest order, we can replace the value $\Delta V_{1}$
with $\Delta V$, since the difference is of higher
order. 
Thus, we can write
\begin{equation}
	N_{0;\mu}^{\mu}=\Delta V/\sigma_{0}.\label{eq:eomgral0}
\end{equation}
For the first correction to the normal vector, Eq.~(\ref{eq:K1}) gives an equation of
the same form, namely, $N_{1;\mu}^{\mu}=$ constant. 
Alternatively, we can obtain an equation for the full $N^{\mu}$ to this order.
Indeed, if we add Eqs.~(\ref{eom1}) and (\ref{eom2}), 
the expression on the left-hand side is the expansion of
$\sigma K|_{0}$, while on the right-hand side we have the expansion
of $\Delta V$, both to next-to-leading order (NLO) in $l/L$. Therefore,
if we only need results to this order, we may write $K|_0=-\Delta V/\sigma$,
where $\sigma=\sigma_{0}+\sigma_{1}$.
Thus, the equation is the same as the lowest-order one, only with the parameter
$\sigma$ updated,
\begin{equation}
N^{\mu}_{\ ;\mu}=\Delta V/\sigma. \label{eq:eomgral1}
\end{equation}

Similarly, from Eq.~(\ref{eq:preK2})
we obtain an equation for $N_{2}^{\mu}$, but we can also add Eqs.~(\ref{eom1})-(\ref{eom3}) 
and identify the expansions of $\sigma K|_{n=0}$
and $\Delta V$ up to the next-to-next-to-leading order (NNLO), 
to obtain an equation for the full $N^\mu$ up to this order,
\begin{equation}
	\sigma K|_{n=0}=-{\Delta V}-({\mu_{0}}/{2})\partial_{n}^{2}K_{0}|_{n=0}. \label{eq:Khasta2}
\end{equation}
We see that the wall EOM is no longer of the form ${N^{\mu}}_{;\mu}=\Delta V/\sigma$.
Moreover, the quantity $\sigma$ is not a constant at this order, 
and this fact should be taken into account explicitly in the equation for $N^{\mu}$. 
Using the separation (\ref{eq:sigma2sep}) and defining the parameter $\tilde{\sigma}=\sigma_{0}+\sigma_{1}+\tilde{\sigma}_{2}$,
we may write Eq.~(\ref{eq:Khasta2}) as
\begin{equation}
	\tilde{\sigma}K|_{n=0}- \mu_0 (K_0 \partial_{n}K_0)|_{n=0} +  (\mu_0/2) \partial_{n}^{2}K_0|_{n=0}
	=-\Delta V.\label{eq:eom3completa}
\end{equation}
We could actually replace $K_0$ by $K$ everywhere, since $\mu_0$ is of order $l^2$.
Thus, we would obtain an equation for the full quantity $K(\xi^{a},0)$ to this order.
Then, using Eqs.~(\ref{eq:Kcovar}) for $K$ and its derivatives, we have an equation for $N^{\mu}$. 
However, solving
this equation can be cumbersome since the second and third terms in (\ref{eq:eom3completa}) are
non-linear in $N^{\mu}$. 
In most cases, it will be more pragmatic to just keep $K$ to leading order in the terms proportional to $\mu_0$. 
Moreover, in these terms we can use the approximation $\tilde{\sigma}=\sigma_{0}$.
Using also the LO equation $K_{0}|_{0}=-\Delta V/\sigma_{0}$, we obtain,
in terms of $N^{\mu}$,
\begin{equation}
N^{\mu}_{\ ;\mu}=\frac{\Delta V}{\tilde{\sigma}} + 
\frac{\mu_0}{\sigma_0} \left( \frac{\Delta V}{\sigma_{0}}N^{\mu}_{0;\nu} N^{\nu}_{0;\mu}
- N^{\mu}_{0;\nu} N^{\nu}_{0;\rho} N^{\rho}_{0;\mu} \right).
\label{eq:eomgral2}
\end{equation}
The left-hand side of Eq.~(\ref{eq:eomgral2}) is like in the previous orders, while the right-hand side contains source terms
which depend on the leading-order solution.
The only vestige of the NLO correction is the term $\sigma_{1}$
in the quantity $\tilde{\sigma}$.

For a domain wall (i.e., in the case $\Delta V=0$), the LO and NLO equations are exactly the same, namely, ${N^{\mu}}_{;\mu}=0$, while
the NNLO equation is ${N^{\mu}}_{;\mu}=-(\mu_0/\sigma_0){N_{0}^{\mu}}_{;\nu}{N_{0}^{\nu}}_{;\rho}{N_{0}^{\rho}}_{;\mu}$.
This result is in agreement with  Ref.~\cite{ghg90}, where it was obtained
using an iterative process (see also \cite{gg90,sm93,l93,al94,a95} for similar calculations).
In this case, we have $\tilde{V}=0$ and $K_0|_0=0$, which give $f_1=0$, and, hence, $\phi_1=0$, i.e., there is no correction to the wall profile at this order. 
Since we have $K_1|_0=0$, we obtain $\tilde{f}_2=0$, which implies
$\phi_{2}=\phi_{2b}(n)\partial_{n}K_{0}(\xi^{a},0)$.

To put the wall EOM in terms of the wall position $X^{a}(\xi^{a})$, 
we shall use the Monge parametrization $x^3=x_{w}^{3}(x^{0},x^{1},x^{2})$ 
and replace $N_{\mu}=(\partial_{a}x_{w}^{3},-1)/s$ in the above results.
We can proceed in different ways. 
For instance, writing $x_{w}^{3}=x_{w0}^{3}+x_{w1}^{3}+\cdots$, 
we may obtain an equation for each term $x_{wi}^3$ from Eqs.~(\ref{eq:K0})-(\ref{eq:preK2}) and solve order by order. 
This method will give nonlinear equations due to the normalization factor $1/s$. 
An alternative is to complete an equation for the full 
quantity $N^{\mu}$ up to the highest order under consideration,
as discussed below Eq.~(\ref{eq:eom3completa}).
This gives an equation for a  single function $x_w^3$.
Such an equation will contain non-linear terms, but 
will be particularly useful for treating linearized perturbations  such a small deformations from a planar or spherical wall.
Another possibility is to solve the EOM for $x_w^3$ up to a given order and then use the solution
as a source term in the next-order EOM, 
as in the sequence of equations (\ref{eq:eomgral0}), (\ref{eq:eomgral1}), (\ref{eq:eomgral2}).
Let us consider this recursive approach.

In Ref.~\cite{mm23} we wrote the LO and NLO equations for $x_{w}^{3}$ (which have the same form) without specifying
a particular coordinate system. 
Beyond the next-to-leading order, such general expressions become very cumbersome.
In Minkowski space and using Cartesian coordinates $x^{a}=t,x,y$, $x^{3}=z$ with the metric $\eta_{\mu\nu}=\mathrm{diag}(1,-1,-1,-1)$,
we have the parametrization $z=z_{w}(t,x,y)$, and
the leading-order EOM, Eq.~(\ref{eq:eomgral0}), becomes
\begin{equation}
	\partial_{a}\frac{\partial^{a}z_{w0}}{s_{0}}=\frac{\Delta V}{\sigma_{0}},\label{eq:eom0_Mink}
\end{equation}
with $s_{0}=\sqrt{1-\partial_{c}z_{w0}\partial^{c}z_{w0}}$. The NLO
EOM, Eq.~(\ref{eq:eomgral1}), takes the same form, with the parameter
$\sigma_{0}$ replaced by $\sigma_{0}+\sigma_{1}$. 
Using the relation $l_0^2=\mu_0/\sigma_0$ [from Eq.~(\ref{eq:defl})],
the NNLO EOM,
Eq.~(\ref{eq:eomgral2}), becomes
\begin{equation}
	\partial_{a}\frac{\partial^{a}z_{w}}{s}=\frac{\Delta V}{\tilde{\sigma}}+l_{0}^{2}\frac{\Delta V}{\sigma_{0}}\,\partial_{b}\frac{\partial^{a}z_{w0}}{s_{0}}\partial_{a}\frac{\partial^{b}z_{w0}}{s_{0}}-l_{0}^{2}\,\partial_{b}\frac{\partial^{a}z_{w0}}{s_{0}}\partial_{c}\frac{\partial^{b}z_{w0}}{s_{0}}\partial_{a}\frac{\partial^{c}z_{w0}}{s_{0}},\label{eq:eom2_Mink}
\end{equation}
with $s=\sqrt{1-\partial_{a}z_{w}\partial^{a}z_{w}}$. 
These equations of motion are still general, but are most useful to treat deformations of a planar wall.
In Sec.~\ref{sec:EOM} we consider spherical coordinates
for the case of a spherically symmetric bubble.

Finally, let us consider the surface tension and the wall width.
Using Eq.~(\ref{eq:sigma2sep}), the former can be written in the form $\sigma=\tilde{\sigma}+\delta\sigma(\xi^{a})$,
with
\begin{equation}
	\tilde{\sigma}=\sigma_{0}+\sigma_{1}+\tilde{\sigma}_{2},
	\quad
	\delta\sigma=-\mu_{0}\partial_{n}K_{0}|_{0}, \label{eq:sigmasep}
\end{equation} 
On the other hand, the definition (\ref{eq:defl}) gives 
$ l_{0}=\sqrt{{\mu_{0}}/{\sigma_{0}}}$.
To obtain $l$ to higher orders, we will need the expansion of $\mu$. 
Proceeding like we did for $\sigma$, we obtain
\begin{equation}
	\mu_1=\int_{-\infty}^{+\infty}\left(f_1n^2 + 2\phi_0^{\prime 2}C_1 n \right)dn
	\label{mu1}
\end{equation}
and $\mu_{2}=\tilde{\mu}_{2}+\alpha\partial_{n}K_{0}(\xi^{a},0)$, with
\begin{equation}
	\tilde{\mu}_{2}=\int_{-\infty}^{+\infty}
	\big[(\phi_{1}^{\prime2}+\tilde{f}_{2})n^{2}+2\phi_{0}^{\prime 2}C_{2a}n \big]dn, 
	\quad
	\alpha=\int_{-\infty}^{+\infty}\left( 2C_{2b}n -{\textstyle\frac{1}{3}}n^{4} \right) \phi_{0}^{\prime 2} dn .
	\label{eq:sigmamu2}
\end{equation}
Therefore, Eq.~(\ref{eq:defl}) gives $l=\tilde{l}+\delta l(\xi^{a})$, with
\begin{align}
	\tilde{l} & =l_{0}\left[1+\frac{1}{2}\frac{\mu_{1}}{\mu_{0}}-\frac{1}{2}\frac{\sigma_{1}}{\sigma_{0}}-\frac{1}{8}\frac{\mu_{1}^{2}}{\mu_{0}^{2}}
	+\frac{3}{8}\frac{\sigma_{1}^{2}}{\sigma_{0}^{2}}-\frac{1}{4}\frac{\sigma_{1}}{\sigma_{0}}\frac{\mu_{1}}{\mu_{0}}
	+\frac{1}{2}\frac{\tilde{\mu}_{2}}{\mu_{0}}-\frac{1}{2}\frac{\tilde{\sigma}_{2}}{\sigma_{0}}\right], \label{expanltil}\\
	\delta l & =\frac{l_{0}}{2}\left[\frac{\alpha}{\mu_{0}}+\frac{\mu_{0}}{\sigma_{0}}\right]\partial_{n}K_{0}|_{0} . \label{expanlvar}
\end{align}
These quantities may depend on the point of the hypersurface through the quantity $\partial_{n}K_{0}$, which is given by Eq.~(\ref{eq:Kcovar}). We have
\begin{equation}
	\partial_{n}K_{0}|_0=\partial_{b}\frac{\partial^{a}z_{w0}}{s_{0}}\partial_{a}\frac{\partial^{b}z_{w0}}{s_{0}}.\label{eq:dnK0_Mink}
\end{equation}
This quantity also appears in the expression for the field profile,
Eq.~(\ref{eq:fitot}). The latter depends also on the variable $n$, which is given by Eq.~(\ref{eq:n_orden2}).
We have
\begin{multline}
	n=\frac{z-z_{w}}{s}-\frac{\partial^{c}z_{w}\partial^{b}z_{w}\partial_{b}\partial_{c}z_{w}}{2s^{3}}\frac{\left(z-z_{w}\right)^{2}}{s^{2}}\\
	+\left[\frac{(\partial^{c}z_{w}\partial^{b}z_{w}\partial_{b}\partial_{c}z_{w})^{2}}{2s^{6}}+\frac{\partial^{a}z_{w}\partial^{b}z_{w}\partial^{c}z_{w}\partial_{b}\partial_{c}\partial_{a}z_{w}}{6s^{4}}\right]\frac{\left(z-z_{w}\right)^{3}}{s^{3}}+\cdots.\label{eq:n_orden2_Monge_Cart}
\end{multline}

\section{The wall motion}

\label{sec:EOM}

We will use Eq.~(\ref{eq:eom2_Mink}) to study the evolution of a  planar wall and deformations thereof.
It is worth emphasizing that this EOM makes no assumptions about the shape of the wall,
and is valid as long as the parametrization $z=z_{w}(t,x,y)$ holds, which is always possible locally. 
A similar but more cumbersome equation can be obtained in spherical coordinates
(see our previous work \cite{mm23} for the leading-order equation), which will be more appropriate 
for a realistic bubble.
For the case of a spherical wall without deformations, we will derive the EOM directly from Eq.~(\ref{eq:eomgral2}).

\subsection{Planar wall}

The evolution of a spatially-planar wall is equivalent to a 1+1 dimensional problem. 
We have $z_{w}(t,x,y)=z_{w}(t)$, $s=\sqrt{1-\dot{z}_{w}^{2}}=\gamma_{w}^{-1}$,
and Eqs.~(\ref{eq:eom0_Mink})-(\ref{eq:eom2_Mink}) become
\begin{align}
\gamma_{w0}^{3}\ddot{z}_{w0} & =\Delta V/\sigma_{0},\label{eq:eompl0}\\
\gamma_{w}^{3}\ddot{z}_{w} & =\frac{\Delta V}{\tilde{\sigma}}+l_{0}^{2}\frac{\Delta V}{\sigma_{0}}\left(\gamma_{w0}^{3}\ddot{z}_{w0}\right)^{2}-l_{0}^{2}\left(\gamma_{w0}^{3}\ddot{z}_{w0}\right)^{3}.\label{eq:eompl2}
\end{align}
Inserting the former in the latter, we obtain
\begin{equation}
\gamma_{w}^{3}\ddot{z}_{w}=\Delta V/\tilde{\sigma}.\label{eq:eompl2b}
\end{equation}
In the case of a domain wall, with $\Delta V=0$, this equation is just $\ddot{z}_{w}=0$, as expected.
For $\Delta V\neq0$, the NNLO equation has the same form as the LO one, with the replacement $\sigma_{0}\to\tilde{\sigma}$.
The EOM also gives the mean curvature of the hypersurface, and in this case we have $K=-\gamma_{w}^{3}\ddot{z}_{w} =$ constant. 
Therefore, we define the parameter $L\equiv |K|^{-1}$ associated to the curvature radius%
\footnote{Since $K=\gamma^{ab}K_{ab}$ is the the trace of the extrinsic curvature tensor, 
	for a three-dimensional hypersurface we have $K=k_{1}+k_{2}+k_{3}$, where the eigenvalues $k_{i}$ give the principal curvatures \cite{spivak3,spivak4}, and we may define the corresponding curvature radii $L_i = |k_i|^{-1}$.
	For a spatially planar wall, we have $K_{\mu\nu}=
-(\delta_{\mu}^{0}\delta_{\nu}^{0}-\delta_{\mu}^{0}\delta_{\nu}^{z}\dot{z}_{w}-\delta_{\mu}^{z}\delta_{\nu}^{0}\dot{z}_{w}
+\delta_{\mu}^{z}\delta_{\nu}^{z}\dot{z}_{w}^{2}) \gamma_{w}^{5}\ddot{z}_{w}$, and
there is only one non-vanishing eigenvalue, $k_1=-\gamma_{w}^{3}\ddot{z}_{w}$, with eigenvector $\gamma_w(1,0,0,-\dot{z}_w)$.}. 
At leading order we have $L_{0}=\sigma_{0}/\Delta V$
and at next-to-next-to-leading order we have $L=\tilde{\sigma}/\Delta V$. 
We remark that $\tilde{\sigma}$ is not the total NNLO surface
tension, which is given by Eq.~(\ref{eq:sigmasep}) as $\sigma=\tilde{\sigma}+\delta\sigma$. 
In the present case, we have $\delta\sigma=-\sigma_{0}l_{0}^{2}/L_{0}^{2}$, but this contribution disappears from $K$
as the last two terms in Eq.~(\ref{eq:eompl2}) cancel each other out.

The leading-order EOM, Eq.~(\ref{eq:eompl0}), has the first integral
$\gamma_{w0}-\gamma_{i0}=(z_{w0}-z_{i0})/L_{0}$, where $z_{i0},\gamma_{i0}$
are the initial values. It is straightforward to solve algebraically
for $\dot{z}_{w0}$ and then integrate the first-order differential
equation. We obtain
\begin{equation}
z_{w0}=z_{i0}-L_{0}\gamma_{i0}+\sqrt{L_{0}^{2}+(t+L_{0}\gamma_{i0}\dot{z}_{i0})^{2}}.\label{eq:solpl0}
\end{equation}
Similarly, we have $\gamma_{w}=\gamma_{i}+(z_{w}-z_{i})/L$ and
\begin{equation}
z_{w}=z_{i}-L\gamma_{i}+\sqrt{L^{2}+(t+L\gamma_{i}\dot{z}_{i})^{2}} . \label{eq:solpl2}
\end{equation}
The initial conditions depend on the problem under consideration and
can be different at each order. In the specific case of a vacuum phase
transition in 1+1 dimensions, $z_{i}$ corresponds to the bubble radius
at nucleation. For consistency, in that case $z_{i}$ should be calculated
using the thin-wall approximation for the instanton to the corresponding
order. We discuss such a calculation in Sec.~\ref{sec:nucleation}.
On the other hand, if we just want to describe the evolution of a
planar wall from a given initial condition, then we should use the same
value of $z_{i}$ at each order. 

Writing Eq.~(\ref{eq:solpl2})
in the form $(z_{w}-z_{i}+L\gamma_{i})^{2}=L^{2}+(t+L\gamma_{i}\dot{z}_{i})^{2}$,
we see that the solution is invariant under Lorentz transformations. 
This is more apparent if we translate the origin of the coordinate system to the point $t_{0}=-L\gamma_{i}\dot{z}_{i}$,
$z_{0}=z_{i}-L\gamma_{i}$, so that the new initial values are $z_{i}=L$,
$\dot{z}_{i}=0$ and we have $\gamma_{w}=z_{w}/L$ and $z_{w}^{2}=L^{2}+t^{2}$.
This result is in agreement with the existence of a Lorentz-invariant
solution of Eq.~(\ref{eq:eccampo}) for $\phi(z,t)$.

Using the above expressions in Eqs.~(\ref{eq:dnK0_Mink}) and (\ref{eq:n_orden2_Monge_Cart}), we obtain the quantities
$\partial_{n}K_0|_{0}=(\Delta V/\sigma_{0})^{2}$ and
\begin{equation}
	n=\gamma_{w}(z-z_{w})-\frac{\dot{z}_{w}^{2}}{2}\frac{\Delta V}{\sigma}\left[\gamma_{w}(z-z_{w})\right]^{2}+\mathcal{O}\left((z-z_{w})^{4}\right) \label{eq:n2pl}
\end{equation}
(the NNLO term vanishes in the expansion of $n$).
To leading order, the coordinate transformation (\ref{eq:n2pl}) is
essentially a Lorentz boost to a coordinate system instantaneously
at rest with the wall. 

\subsection{Small deformations}

Let us now consider small deformations of a planar wall.
For an infinitely thin wall, we write 
$z_{w0}=z_{\mathrm{pl}0}(t)+\delta z_{0}(t,x,y)$,
where $z_{\mathrm{pl}0}$ is a solution of the planar-wall EOM, Eq.~(\ref{eq:eompl0}), and $\delta z_{0}$ is a small perturbation.
Keeping terms up to linear order in $\delta z_{0}$ in  Eq.~(\ref{eq:eom0_Mink}),
we obtain the equation
\begin{equation}
	\partial_{t}(\gamma_{\mathrm{pl}0}^{3}\partial_{t}\delta z_{0})-\gamma_{\mathrm{pl}0}(\partial_{x}^{2}\delta z_{0}+\partial_{y}^{2}\delta z_{0})=0.\label{eq:eompl_lin0}
\end{equation}
To go beyond the zero-thickness approximation for both the background solution and the perturbation,
we write $z_{w}=z_{\mathrm{pl}}(t)+\delta z(t,x,y)$,
where $z_{\mathrm{pl}}$ fulfills the NNLO planar wall EOM, Eq.~(\ref{eq:eompl2b}), and we
keep terms up to linear order in $\delta z$ in Eq.~(\ref{eq:eom2_Mink}).
We obtain
\begin{equation}
	\partial_{t}\left(\gamma_{\mathrm{pl}}^{3}\partial_{t}\delta z\right)-\gamma_{\mathrm{pl}}\left(\partial_{x}^{2}+\partial_{y}^{2}\right)\delta z=-\frac{l_{0}^{2}}{L_{0}^{2}}\gamma_{\mathrm{pl}0}\left(\partial_{x}^{2}+\partial_{y}^{2}\right)\delta z_{0}.\label{eq:eompl_lin2}
\end{equation}
With the convenient change of variables $dt=\gamma_{\mathrm{pl}0}d\tau_{0}$, $\delta z_{0}=\xi_{0}/\gamma_{\mathrm{pl}0}$,
Eq.~(\ref{eq:eompl_lin0}) becomes
\begin{equation}
	\left(\partial_{\tau_{0}}^{2}-\partial_{x}^{2}-\partial_{y}^{2}\right)\xi_{0}-L_{0}^{-2}\xi_{0}=0. \label{eq:econdas}
\end{equation}
This equation was previously derived in Ref.~\cite{gv91} (for the case $\dot{z}_{i0}=0$) using a covariant gauge.
From Eq.~(\ref{eq:solpl0}), we have $\gamma_{\mathrm{pl}0}=\sqrt{1+(t/L_{0}+\gamma_{i0}\dot{z}_{i0})^{2}}$ and we obtain
\begin{equation}
	\tau_{0}= L_{0}\left[\mathrm{arcsinh}(t/L_{0}+\gamma_{i0}\dot{z}_{i0})-\mathrm{arcsinh}(\gamma_{i0}\dot{z}_{i0})\right] .
\end{equation}
Similarly, with the change of variables $dt=\gamma_{\mathrm{pl}}d\tau$, $\delta z=\xi/\gamma_{\mathrm{pl}}$,
Eq.~(\ref{eq:eompl_lin2}) becomes 
\begin{equation}
	\left(\partial_{\tau}^{2}-\partial_{x}^{2}-\partial_{y}^{2}\right)\xi -L^{-2}\xi =
	-(l_{0}/L_{0})^2\left(\partial_{x}^{2}+\partial_{y}^{2}\right)\xi_{0}. \label{eq:econdasNNLO}
\end{equation}
From Eq.~(\ref{eq:solpl2}), we have  $\gamma_{\mathrm{pl}}=\sqrt{1+(t/L+\gamma_{i}\dot{z}_{i})^{2}}$ and we obtain
\begin{equation}
	\tau=L\left[\mathrm{arcsinh}(t/L+\gamma_{i}\dot{z}_{i})-\mathrm{arcsinh}(\gamma_{i}\dot{z}_{i})\right] .
\end{equation}

The LO equation (\ref{eq:econdas}) has solutions of the
form $\xi_{0}=A_{0}e^{i(\mathbf{k}\cdot\mathbf{x}\pm\omega_{0}\tau_{0})}$,
where $\mathbf{x}=(x,y)$, $\mathbf{k}=(k_{x},k_{y})$, and $\omega_{0}=\sqrt{k^{2}-L_{0}^{-2}}$,
with $k^{2}=k_{x}^{2}+k_{y}^{2}$. As noticed in \cite{gv91}, although
this solution for $\xi_{0}$ grows exponentially with $\tau_{0}$
for $k^{2}<L_{0}^{-2}$, these modes do not have an exponential growth
in the variable $t$, and, furthermore, the gamma factor in $\delta z_{0}=\xi_{0}/\gamma_{\mathrm{pl}0}$
causes $\delta z_{0}$ to actually decrease with $t$.%
\footnote{Taking $\dot{z}_{i0}=0$ for simplicity, we have 
	$e^{\pm i\omega_{0}\tau_{0}}=(t/L_{0}+\gamma_{\mathrm{pl}0})^{\pm\sqrt{1-(kL_{0})^{2}}}$,
	with $\gamma_{\mathrm{pl}0}=\sqrt{1+(t/L_{0})^{2}}$. 
	Therefore, we have $e^{\pm i\omega_{0}\tau_{0}}/ \gamma_{\mathrm{pl}0} \sim t^{\pm\sqrt{1-(kL_{0})^{2}}-1}$ for $t\to\infty$.} 
For the other modes, $\xi_{0}$ oscillates with frequency $\omega_{0}$
with respect to $\tau_{0}$. With respect to $t$, these oscillations
become increasingly slower, and, besides, $\delta z_{0}$ decreases
as $\gamma_{\mathrm{pl}0}^{-1}$. Let us consider for concreteness
the initial conditions $\dot{z}_{i0}=0$ and $\partial_{t}\delta z_{0}=0$.
We have
\begin{equation}
	\delta z_{0}= 
	A_{0}\frac{\cos\left[\omega_{0}L_{0}\,\mathrm{arcsinh}(t/L_{0})\right]}{\sqrt{1+(t/L_{0})^{2}}}
	\cos(\mathbf{k}\cdot\mathbf{x}).
	\label{eq:deltazk0}
\end{equation}

For the NNLO equation (\ref{eq:econdasNNLO}), we may propose a solution of the form $A(\tau)e^{i\mathbf{k}\cdot\mathbf{x}}$.
The equation for $A(\tau)$ is that of a harmonic oscillator subjected
to an external force, which can be solved analytically. However, a
simpler approach in this case is to just use the approximation $\xi_{0}=\xi$
in the NNLO term.
This approximation will be valid as long as both $l_{0}/L_{0}$ and $l_{0}k$ are small, since the wall has a local spatial curvature of order $k$.
The solutions are of the form $\xi=Ae^{i(\mathbf{k}\cdot\mathbf{x}\pm\omega\tau)}$,
with $\omega=\sqrt{(1-l_{0}^{2}/L_{0}^{2})k^{2}-L^{-2}}$.
For the initial conditions $\dot{z}_{i}=0$ and $\partial_{t}\delta z=0$, we have
\begin{equation}
	\delta z=
	A\frac{\cos\left[\omega L\,\mathrm{arcsinh}(t/L)\right]}{\sqrt{1+(t/L)^{2}}}\cos(\mathbf{k}\cdot\mathbf{x}). \label{eq:deltazk2}
\end{equation}

According to Eqs.~(\ref{eq:sigmasep})-(\ref{expanlvar}), the
variations of the surface will induce variations of $\sigma$ and
$l$ which depend on the quantity $\partial_nK_0$.
Using Eqs.~(\ref{eq:eompl0}) and (\ref{eq:eompl_lin0}) in Eq.~(\ref{eq:dnK0_Mink}), we have 
\begin{equation}
	\partial_{n}K_{0}|_0=
	L_{0}^{-2}+2L_{0}^{-1}\gamma_{\mathrm{pl}0} (\partial_{x}^{2}+\partial_{y}^{2})\delta z_{0}.\label{eq:dnK0plpert}
\end{equation}
For the solution (\ref{eq:deltazk0}), we have
\begin{equation}
	\partial_{n}K_{0}|_0=
	L_{0}^{-2}-2L_{0}^{-1}A_{0}k^{2}\cos\left[\omega_{0}L_{0}\,\mathrm{arcsinh}(t/L_{0})\right]\cos(\mathbf{k}\cdot\mathbf{x}).
	\label{eq:dnK0k}
\end{equation}
The constant term in Eq.~(\ref{eq:dnK0k}) is already present for
a planar wall and is due to the global acceleration. 
Since we have $\delta l \propto \partial_{n}K_{0}|_0$, we see how 
the spatial undulations of the surface are transmitted to the wall width.
According to Eq.~(\ref{eq:dnK0k}),
these variations do not decrease with time in the wall frame.
However, for an observer at rest at $z=0$, they do decrease due to the Lorentz contraction.

\subsection{Spherical wall}

In spherical coordinates
$x^{a}=\xi^{a}=t,\theta,\varphi$ and $x^{3}=r$, we have a diagonal
metric given by $g_{00}=1,g_{rr}=-1$, $g_{\theta\theta}=-r^{2}$,
$g_{\varphi\varphi}=-r^{2}\sin^{2}\theta$. 
In the Monge representation, the position of the wall is described by $r=r_{w}(t,\theta,\varphi)$.
We will consider
for simplicity the case of a spherical wall,
$r=r_{w}(t)$. 
Thus, Eq.~(\ref{eq:Nmu}) gives $N_{t}=\gamma_{w}\dot{r}_{w}$, $N_{r}=-\gamma_{w}$, 
with $\gamma_{w}=1/\sqrt{1-\dot{r}_{w}^{2}}$, and the other components of the normal vector
vanish. 
The only non-zero Christoffel symbols that appear in Eq.~(\ref{eq:Kcovar}) are $\Gamma_{r\theta}^{\theta}=r^{-1}$ and
$\Gamma_{r\varphi}^{\varphi}=r^{-1}$, 
and Eqs.~(\ref{eq:eomgral0}) and (\ref{eq:eomgral2}) give
\begin{align}
\gamma_{w0}^{3}\ddot{r}_{w0}+2\frac{\gamma_{w0}}{r_{w0}} & =\frac{\Delta V}{\sigma_{0}},\label{eom0esf}\\
\gamma_{w}^{3}\ddot{r}_{w}+\frac{2\gamma_{w}}{r_{w}} & =\frac{\Delta V}{\tilde{\sigma}}+l_{0}^{2}\frac{\Delta V}{\sigma_{0}}\left[\left(\gamma_{w0}^{3}\ddot{r}_{w0}\right)^{2}+2\frac{\gamma_{w0}^{2}}{r_{w0}^{2}}\right]-l_{0}^{2}\left[\left(\gamma_{w0}^{3}\ddot{r}_{w0}\right)^{3}+2\frac{\gamma_{w0}^{3}}{r_{w0}^{3}}\right], \label{eom2esf_0}
\end{align}
while the NLO equation takes the same form as the LO equation, with the replacement $\sigma_{0}\to\sigma_{0}+\sigma_{1}$.
Using (\ref{eom0esf}) in (\ref{eom2esf_0}), we obtain
\begin{equation}
\gamma_{w}^{3}\ddot{r}_{w}+\frac{2\gamma_{w}}{r_{w}}=\frac{\Delta V}{\tilde{\sigma}}+2l_{0}^{2}\frac{\gamma_{w0}}{r_{w0}}\left[\left(\frac{\Delta V}{\sigma_{0}}\right)^{2}-3\frac{\Delta V}{\sigma_{0}}\frac{\gamma_{w0}}{r_{w0}}+3\left(\frac{\gamma_{w0}}{r_{w0}}\right)^{2}\right].\label{eom2esf}
\end{equation}

The case $\Delta V=0$ (domain wall), has already been considered in Ref.~\cite{ghg90}.
At leading order, the wall collapses due to the surface tension.
The NNLO correction, which is proportional to $r_{w0}^{-3}\gamma_{w0}^{3}$, slows down the collapse
and eventually prevents it. However, for small enough $r_{w}$, the
perturbative expansion will break down.

For $\Delta V\neq0$, the first term on the right-hand side of Eq.~(\ref{eom2esf}) is like at leading order,
with the replacement $\sigma_0\to\tilde{\sigma}$. 
It is not difficult to see that the term proportional to $l_0^2$ is always positive.
The case $\Delta V<0$ corresponds to a false-vacuum bubble and, for a vacuum phase transition, has
been studied mainly in curved spacetime (see, e.g., \cite{bgg87,aps89,aj05,lllnp08,nw11}). 
In flat space, such a bubble will collapse.
This case may apply to a thermal phase transition with subcritical bubbles \cite{gkw91,gg94,ghk97}
or to the evolution of false-vacuum domains after bubble percolation \cite{w84,gr98}.
Like in the domain wall case, the correction
proportional to $l_{0}^{2}$ in Eq.~(\ref{eom2esf}) opposes the collapse of the bubble.
For $\Delta V>0$, the bubble
can expand or collapse, depending on whether the pressure difference
or the surface tension dominates. 
This is easily seen in the leading order equation (\ref{eom0esf}), where we have a critical
radius, $r_{c0}=2\sigma_{0}/\Delta V$, for which a bubble with $\dot{r}_{w0}=0$
is in unstable equilibrium.
This case may be relevant for 
the dynamics of phase mixing in phase transitions with subcritical bubbles.
The NNLO corrections shift the critical radius and slow down the bubble collapse. 
We shall focus here on the case of an expanding bubble and address the other possibilities elsewhere.

We define the parameter $R_{0}=3L_{0}=3\sigma_{0}/\Delta V$, which
is the relevant length scale associated to the curvature in the spherical case%
\footnote{Like in the planar case, we have a constant mean curvature $K_{0}|_0=\Delta V/\sigma_{0}=L_{0}^{-1}$.
	However, in the spherical case we have three non-vanishing eigenvalues $k_{i}$
	(associated to the space curvature and the acceleration). If the hypersurface
	is O(3,1) invariant, these eigenvalues are equal and the trace is
	$K_{0}=3k_{i}$. This gives $k_{i}=K_{0}/3$ and $R_{0}=3L_{0}$.
	In Sec.~\ref{sec:nucleation} we discuss this symmetric case.}. 
For convenience of notation, we also define the parameter $\tilde{R}=3\tilde{\sigma}/\Delta V$. 
The latter is not directly related to the curvature, since the
term proportional to $l_0^2$ in Eq.~(\ref{eom2esf}) never vanishes [cf.~Eqs.~(\ref{eq:Khasta2})-(\ref{eq:eom3completa})].

To solve Eq.~(\ref{eom0esf}), we multiply it by $r_{w0}^{2}\dot{r}_{w0}$
and integrate. With the initial conditions $r_{w0}=r_{i0}$, $\gamma_{w0}=\gamma_{i0}$
at $t=0$, we obtain
\begin{equation}
\gamma_{w0}r_{w0}^{2}-\gamma_{i0}r_{i0}^{2}=R_{0}^{-1}\left(r_{w0}^{3}-r_{i0}^{3}\right).\label{eq:gammaw0}
\end{equation}
Solving for $\dot{r}_{w0}$ and integrating again, we obtain the implicit solution
\begin{equation}
\int_{r_{i0}}^{r_{w0}}\frac{dr}{\sqrt{1-(r/R_{0}-\beta_{0}r_{i0}^{2}/r^{2})^{-2}}}=t,\label{eq:solrw0}
\end{equation}
where $\beta_{0}=r_{i0}/R_{0}-\gamma_{i0}$. 
Similarly, we multiply Eq.~(\ref{eom2esf}) by $r_{w}^{2}\dot{r}_{w}$
(taking into account that, in the highest-order term, we have $r_{w}\simeq r_{w0}$)
and integrate. We obtain
\begin{equation}
\gamma_{w}r_{w}^{2}-\gamma_{i}r_{i}^{2}=\tilde{R}^{-1}\left(r_{w}^{3}-r_{i}^{3}\right)+l_{0}^{2}\left[2R_{0}^{3}\left(r_{w0}^{3}-r_{i0}^{3}\right)+\beta_{0}^{3}r_{i0}^{6}\left(r_{w0}^{-6}-r_{i0}^{-6}\right)\right],
\label{eq:solcoleman2-3d}
\end{equation}
where $r=r_i$, $\gamma=\gamma_i$ are the initial conditions at $t=0$.
Solving for $\dot{r}_{w}$ and integrating again (and neglecting orders
higher than $l_{0}^{2}$), we obtain
\begin{equation}
\int_{r_{i}}^{r_{w}} \frac{dr}{\sqrt{1- (r/\tilde{R} - \beta r_{i}^{2}/r^{2} )^{-2}}} = t + 
\frac{l_{0}^{2}}{R_{0}^{2}} \int_{r_{i0}}^{r_{w0}}
\frac{ 
	\frac{2r}{R_{0}} - \left(\frac{2r_{i0}^{3}}{R_{0}^{3}}+\beta_{0}^{3}\right) \frac{R_{0}^{2}}{r^{2}}
	+ \frac{R_{0}^{2}\beta_{0}^{3}r_{i0}^{6}}{r^{8}}
	}{\left[(r/R_{0}-\beta_{0}r_{i0}^{2}/r^{2})^{2}-1\right]^{3/2}} dr, \label{eq:solrw2}
\end{equation}
where $\beta=r_{i}/\tilde{R}-\gamma_{i}$.

At leading order, 
the critical radius between expansion and collapse is $r_{c0}=2L_{0}$. 
Another special situation occurs when the initial conditions fulfill $r_{i0}=\gamma_{i0}R_{0}$, 
so that we have $\beta_{0}=0$
(this condition requires $r_{i0}\geq3L_{0}>r_{c0}$). 
In this case, Eq.~(\ref{eq:gammaw0})
becomes $\gamma_{w0}=r_{w0}/R_{0}$, i.e., the initial relation holds
at all time, and the solution (\ref{eq:solrw0}) takes the same form
as in the 1+1-dimensional case (\ref{eq:solpl0}), 
$r_{w0}^{2}=R_{0}^{2}+(t+R_{0}\gamma_{i0}\dot{r}_{i0})^{2}$.
With the additional initial condition $\dot{r}_{i0}=0$,
we obtain the well-known O(3,1)-invariant solution 
$r_{w0}^{2}=R_{0}^{2}+t^{2}$.
The values $\dot{r}_{i0}=0$, $r_{i0}=R_{0}$ are well-motivated initial
conditions for the nucleated bubble since they are obtained in the
thin-wall approximation for the instanton \cite{c77}. The NNLO
equation, Eq.~(\ref{eq:solrw2}), becomes simpler%
\footnote{Neglecting terms of order higher than $l_{0}^{2}$, we obtain $r_{w}=\sqrt{\tilde{R}^{2}+t^{2}}-\frac{2l_{0}^{2}}{R_{0}^{2}}\left(\frac{R_{0}^{3}}{r_{w0}^{3}}+\frac{R_{0}^{2}}{r_{w0}^{2}}-2\frac{R_{0}}{r_{w0}}\right)$.\label{fn:NNLOanalit}}
for $\dot{r}_{i}=0$
and $r_{i}=\tilde{R}$. 
However, there is no actual motivation for this initial condition,
since the resulting solution is not O(3,1) invariant. Neither does the value $\tilde{R}=3\tilde{\sigma}/\Delta V$
give the best estimate for the nucleation radius, which we address in Sec.~\ref{sec:nucleation}. 

Using Eqs.~(\ref{eom0esf}) and (\ref{eq:gammaw0}) in Eq.~(\ref{eq:Kcovar}), we obtain 
\begin{equation}
	\partial_{n}K_{0}|_{n=0}=\left(\frac{1}{L_{0}} -2\frac{\gamma_{w0}}{r_{w0}}\right)^{2}
	+2\frac{\gamma_{w0}^{2}}{r_{w0}^{2}} = 3R_{0}^{-2}+6\beta_{0}^{2}r_{i0}^{4}r_{w0}^{-6}.
	\label{eq:dnKesf}
\end{equation}
On the other hand,
all the Christoffel symbols that appear in Eq.~(\ref{eq:n_orden2}) vanish for the spherical case.
Using the above results, we obtain
\begin{equation}
	n= 
	\gamma_{w}(r-r_{w})
	-\frac{\dot{r}_{w}^{2}}{2} \left(\frac{3}{\tilde{R}}-\frac{2\gamma_{w}}{r_{w}}\right) \left[\gamma_{w}(r-r_{w})\right]^{2}
	- \frac{\dot{r}_{w}^{4}\gamma_{w}}{r_{w}} \left(\frac{1}{R_0} -\frac{\gamma_{w}}{r_{w}}\right)
	\left[\gamma_{w}(r-r_{w})\right]^{3} .
	\label{eq:nesf}
\end{equation}

\section{Specific examples }

\label{sec:ejemplos}

\begin{figure}[tb]
	\includegraphics[width=0.45\textwidth]{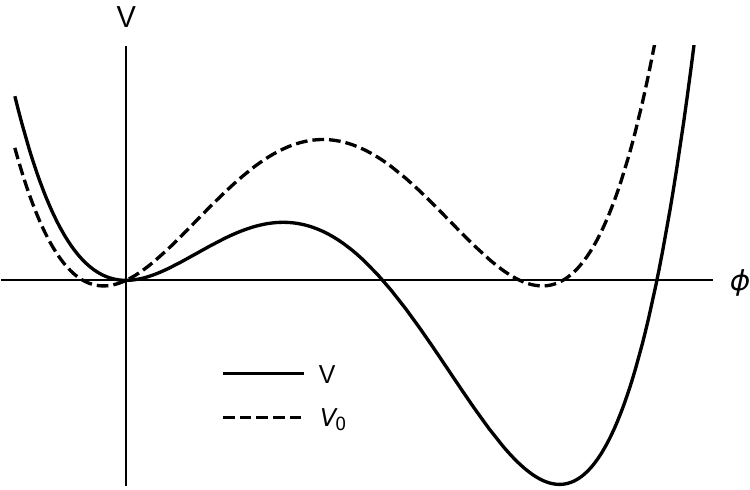}
	\hfill
	\includegraphics[width=0.45\textwidth]{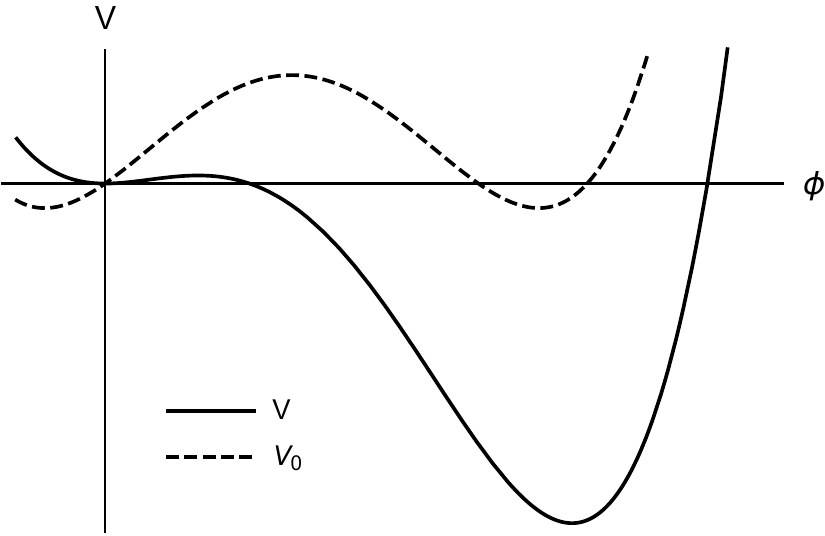}
	\caption{The shape of the potentials (\ref{eq:pot}) and (\ref{eq:V0}).
		Left panel: $\lambda=0.5$, $e/m=1.6$, and $\Delta V/V_{b}\simeq3.5$. 
		Right panel: $\lambda=0.5$, $e/m=1.9$, and $\Delta V/V_{b}\simeq42$.\label{fig:pot}}
\end{figure}

To illustrate our results, we will now consider the simple potential
\begin{equation}
	V(\phi)=\frac{1}{2}m^{2}\phi^{2}-\frac{e}{3}\phi^{3}+\frac{\lambda}{4}\phi^{4},\label{eq:pot}
\end{equation}
where we have $\phi_{+}=0$,
$\phi_{-}=(e/2\lambda)(1+\sqrt{1-4\lambda m^{2}/e^{2}})$,
and the maximum between the two minima is at $\phi_b=e/\lambda-\phi_{-}$.
For $\lambda m^{2}/e^{2}<2/9$, we have $V_{+}=0$,
$V_{-}=-\frac{1}{4}\phi_{-}^{2}(\frac{e}{3}\phi_{-}-m^{2})<0$,
and $V_b\equiv V(\phi_b)=\frac{1}{4}\phi_{b}^{2}(m^{2}-\frac{e}{3}\phi_{b})$.
We will construct the degenerate
potential $V_{0}$ with a linear term $\tilde{V}(\phi)=c\phi$. 
The conditions (\ref{eq:condminV0})-(\ref{eq:condV0deg}) give $c=e(9\lambda m^{2}-2e^{2})/27\lambda^{2}$,
and we have
\begin{equation}
	V_{0}(\phi)-V_{0\pm} =\frac{\lambda}{4}\left(\phi-a_{+}\right)^{2}\left(\phi-a_{-}\right)^{2},\label{eq:V0}
\end{equation}
with $a_{\pm}=\frac{e}{3\lambda}(1\mp\sqrt{1-27c\lambda^{2}/e^{3}})$.
As discussed in Sec.~\ref{apbasica},
the thin-wall approximation is expected to work well when $\Delta V/V_{b}\ll1$.
In our previous work \cite{mm23}, we showed  that the LO  and NLO approximations work quite
well even when we have $\Delta V/V_{b}\sim 1$. Here, we will
consider two parameter sets which give $\Delta V/V_{b}\simeq3.5$ and
$\Delta V/V_{b}\simeq42$ (see Fig.~\ref{fig:pot}).

\subsection{Wall profile and EOM parameters}

The functions of $n$ obtained in Sec.~\ref{sec:pert_method} can be written as functions of $\phi_0$
by means of Eq.~(\ref{eq:fi0}),
and the integrals with respect to $n$ can be written as integrals with respect to the field, like in Eq.~(\ref{eq:nast}).
This simplifies the calculations by avoiding the explicit use of the solution $\phi_0(n)$.
We write down the expressions in App.~\ref{sec:perfil}.
For this specific potential, we obtain analytic results for most of the quantities, which we also write down in App.~\ref{sec:perfil}.
In particular, we have
\begin{equation}
\phi_{0}(n)=\phi_{*}-a\tanh \big(\sqrt{\lambda/2}\,a\,n \big),\label{eq:fi0pot}
\end{equation}
where $\phi_{*}=(a_{+}+a_{-})/2$ and $a=(a_{-}-a_{+})/2$.
The LO wall width is given by
\begin{equation}
l_{0}=\sqrt{(\pi^{2}/6-1)}\,(\sqrt{\lambda} a)^{-1}.\label{eq:l0pot}
\end{equation}
For the potential on the left panel of Fig.~\ref{fig:pot} we have
$l_{0}/L_{0}\simeq0.3$, while for that on the right panel, we have $l_{0}/L_{0}\simeq0.54$.

With the linear modification $\tilde{V}=c\phi$, we obtain 
\begin{equation}
\phi_{1}=-c/(2a^{2}\lambda).\label{eq:fi1pot}
\end{equation}
Since $\phi_{1}$ is a constant, according to Eq.~(\ref{eq:Gi}), the NLO quantities $I_1^{(k)}$, $\bar{I}_1^{(k)}$ vanish.
In particular, we have $\sigma_{1}=\mu_{1}=0$, so, to next-to-leading order, we still
have $l=l_{0}$. 
As discussed in \cite{mm23}, these characteristics of the NLO corrections are a consequence of the
simplicity of the quartic potential and the linear modification, and
do not occur for other potentials or modifications.
Nevertheless, the overall quantitative results are similar.
For example, although the correction $\sigma_1$ vanishes in this case, 
the approximation $\sigma_0$ is already as good as $\sigma_0+\sigma_1$
for other potentials or modifications.

For the NNLO quantities, we also obtain analytic expressions 
(which can be found in App.~\ref{sec:perfil}),
except for the parameter $\alpha$.
The constants $c_{2a}$ and $c_{2b}$ vanish in this case, so we have 
\begin{equation}
\phi_{2}=\phi_{h}(\phi_{0})\left[C_{2a}(\phi_{0})+C_{2b}(\phi_{0})\partial_{n}K_{0}|_0\right].
\label{eq:fi2pot}
\end{equation}
Using the results for $\sigma_i$ and $\mu_i$ in Eqs.~(\ref{expanltil})-(\ref{expanlvar}), we obtain
\begin{equation}
	\frac{l}{l_{0}}=1+\frac{3c^{2}}{8\lambda^{2}a^{6}}
	+\frac{1}{2}\left(\frac{\pi^{2}}{6}-1+\frac{0.346305}{\pi^{2}/6-1}\right)\frac{\partial_{n}K_{0}|_{0}}{\lambda a^{2}} . \label{eq:ltotpot}
\end{equation}
The constant correction 
 proportional to $c^{2}$ comes from the term $\tilde{V}$ which breaks the degeneracy of the potential,
and is due to the wall acceleration.

\subsection{Evolution of a thick spherical bubble}

For a spherical bubble, the LO  radius $r_{w0}$ is given by Eq.~(\ref{eq:solrw0})
and depends on the  parameter $R_{0}=3L_{0}=3\sigma_{0}/\Delta V$
and the initial conditions $r_{i0},\gamma_{i0}$. 
The NLO radius has the same form, with the parameter $\sigma_0$ replaced by $\sigma_0+\sigma_1$.
Moreover, for the specific potentials $V$ and $V_0$ given by Eqs.~(\ref{eq:pot})-(\ref{eq:V0}),
we have $\sigma_1=0$, so, if we take the same initial conditions, we have the same solution $r_{w0}$. 
The NNLO bubble radius is given by Eq.~(\ref{eq:solrw2}) and depends on the parameter $\tilde{R}$ and the initial conditions $r_i,\gamma_i$,
as well as on the LO solution.
On the other hand, 
the total NNLO field profile is given by Eq.~(\ref{eq:fitot}), where
$\partial_{n}K_{0}|_0$ is given by Eq.~(\ref{eq:dnKesf}) 
and the $\phi_{i}$ are given by Eqs.~(\ref{eq:fi0pot}), (\ref{eq:fi1pot}), and (\ref{eq:fi2pot}),
with the variable $n$ given by Eq.~(\ref{eq:nesf}). 

In order to compare these results with a numerical solution of the field equation,
we will solve Eq.~(\ref{eq:eccampo}) for the O(3,1)-symmetric case, which simplifies the numerical computation.
Notice that our perturbative treatment gives an equation of motion for the wall surface
but does not provide initial conditions, i.e., it gives the wall evolution from an initial surface.
In the present case, we will set $r_{i0}=r_i=\bar{r}$, where we will calculate the initial radius $\bar{r}$ from the numerical profile.
For the particular case of the O(3,1)-symmetric solution we could, alternatively, set $r_{i0}=R_{0}$, 
which is the value one obtains by considering the nucleation process in the usual thin-wall approximation \cite{c77}. 
However,  to be consistent with our NNLO EOM, we should use a better estimation for the initial value $r_i$.
We consider this alternative in Sec.~\ref{sec:nucleation}, but here we will just obtain the initial conditions from the numerical solution,
so that we can verify how well our perturbative expansion approximates the subsequent evolution of the wall.

Assuming a solution of Eq.~(\ref{eq:eccampo}) of the form $\phi(x^{\mu})=\bar{\phi}(\rho)$,
with $\rho^2=\mathbf{x}^{2}-t^{2}$, we obtain the well-known equation \cite{c77}
\begin{equation}
\frac{d^{2}\bar{\phi}}{d\rho^{2}}+\frac{3}{\rho}\frac{d\bar{\phi}}{d\rho}=V'(\bar{\phi}) \label{eq:ecColeman}
\end{equation}
with the boundary conditions
\begin{equation}
\frac{d\bar{\phi}}{d\rho}(0)=0,\quad\lim_{\rho\to\infty}\bar{\phi}(\rho)=\phi_{+}.\label{eq:ccColeman}
\end{equation}
This equation is usually solved numerically using the overshoot-undershoot
method. 
Actually, this method
gives only the part of the profile for $r\geq t$ (real $\rho$).
To numerically obtain the profile for $r<t$, Eq.~(\ref{eq:ecColeman})
can be analytically continued, which essentially consists in changing
the sign of the potential. 
The boundary conditions for this part of the profile
are just the matching conditions with the usual solution at $\rho=0$ \cite{mm23,bgm16}. 
We define the wall position $\bar{r}_{w}$ for the numerical
solution as the average of $r$ weighted with the function $(\partial_{r}\phi)^{2}$,
\begin{equation}
	\bar{r}_{w}=\frac{\int_{t}^{\infty}(\partial_{r}\phi)^{2}rdr}{\int_{t}^{\infty}(\partial_{r}\phi)^{2}dr}.\label{eq:rmedio}
\end{equation}
The limit of integration $r=t$ corresponds to $\rho=0$, which can
be regarded as the boundary between the wall and the bubble interior. 

We remark that our perturbative solution for the profile, Eq.~(\ref{eq:fitot}), 
does not take the simple form $\phi(\rho)$ but depends on $x^\mu$ through the quantitites $n$
and $\partial_{n}K_{0}|_0$.
To obtain such a Lorentz-invariant solution in the thin wall approximation,
we must consider the field $\phi$ as a function of $\rho$ from the
beginning. We discuss this approach in Sec.~\ref{sec:nucleation}.
Of course, with the appropriate initial conditions, our general approach will give a good approximation
for this particular case as well, only the symmetry is not explicit
in the perturbative expansion. 
We use the values 
$\gamma_{i0}=\gamma_{i}=1$, $r_{i0}=r_{i}=\bar{r}_{w}(0)$
as initial conditions for our LO and NNLO solutions, Eqs.~(\ref{eq:solrw0}) and (\ref{eq:solrw2}).

\begin{figure}[tb]
\centering
\includegraphics[width=0.49\textwidth]{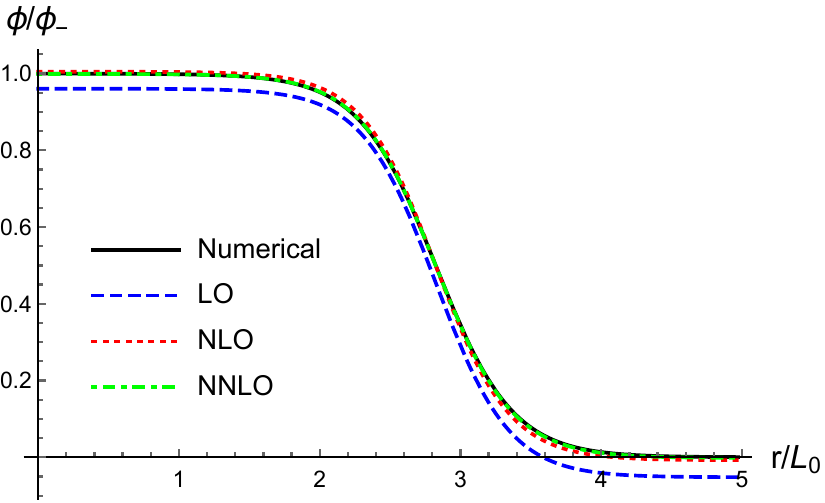}
\hfill
\includegraphics[width=0.49\textwidth]{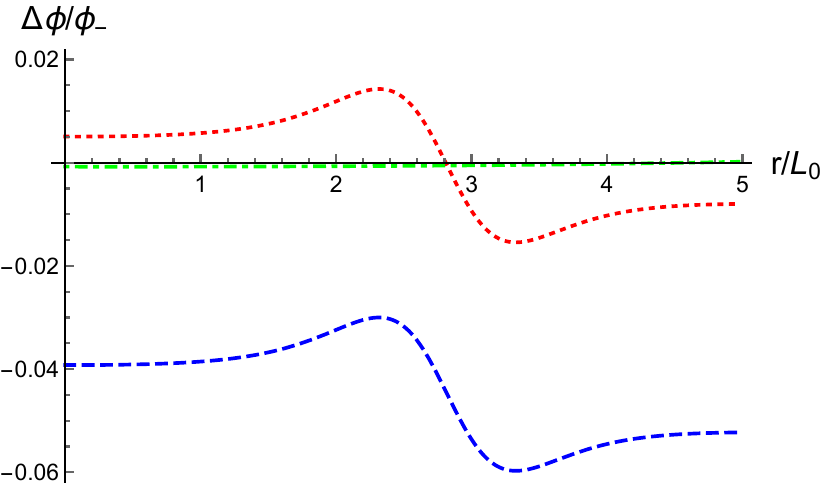}
\caption{Bubble profile at $t=0$ for the potential shape shown in the left
panel of Fig.~\ref{fig:pot} ($\Delta V/V_{b}\simeq3.5$, $l_{0}/L_{0}\simeq0.3$).
Left: The numerical solution and the analytical approximations. Right:
The difference of each approximation with the numerical solution.\label{fig:profg}}
\end{figure}

\begin{figure}[tb]
	\centering
	\includegraphics[height=5.3cm]{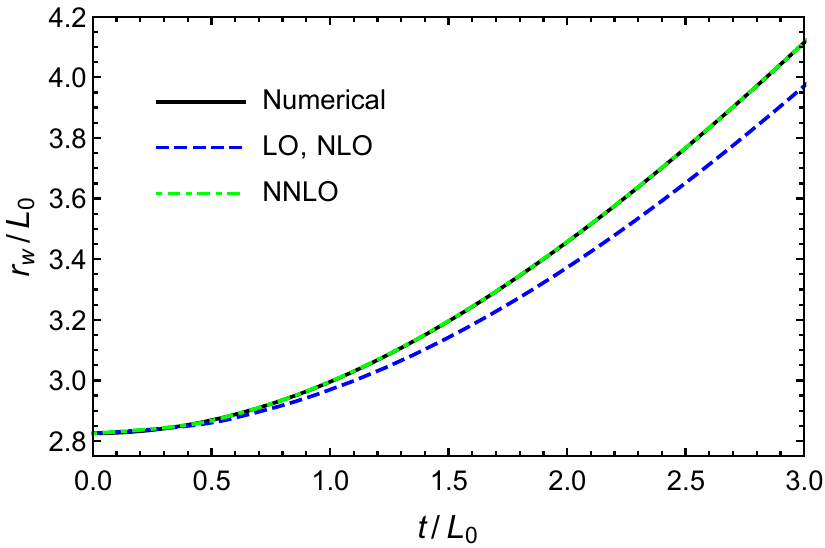}
	\hfill
	\includegraphics[height=5.3cm]{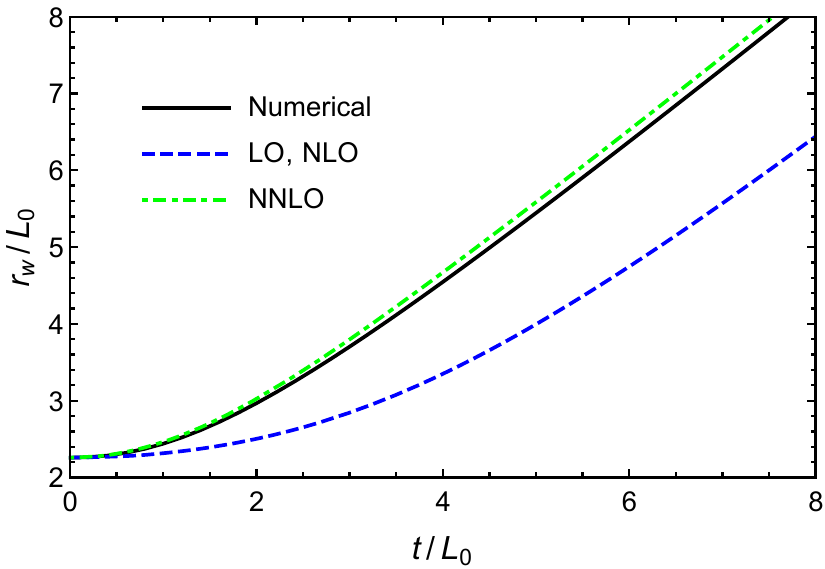}
	\caption{The mean bubble radius for the numerical solution and the
		thin-wall approximations. The left and right panels correspond to
		the potential shapes shown in the left and right panels of Fig.~\ref{fig:pot},
		respectively.\label{fig:rw}}
\end{figure}

In Fig.~\ref{fig:profg}, we consider the case $\Delta V/V_{b}\simeq3.5$.
The field profile is shown at $t=0$, where we have $\rho=r$ and $n=r-r_{i}$.
We see that the NLO approximation is not bad, but the NNLO approximation is much better. 
For more clarity, the right panel shows the difference between the numerical solution and the
approximations.
At later times, the main variation of these profiles will be the change of their position and the 
Lorentz contraction of the wall width.
The different approximations for $r_w$ are shown in the left panel of Fig.~\ref{fig:rw}.
We see that the NNLO approximation is much better than the LO one 
(for this potential, the NLO approximation coincides with the latter since we have $\sigma_1=0$).

\begin{figure}[tb]
\centering
\includegraphics[width=0.49\textwidth]{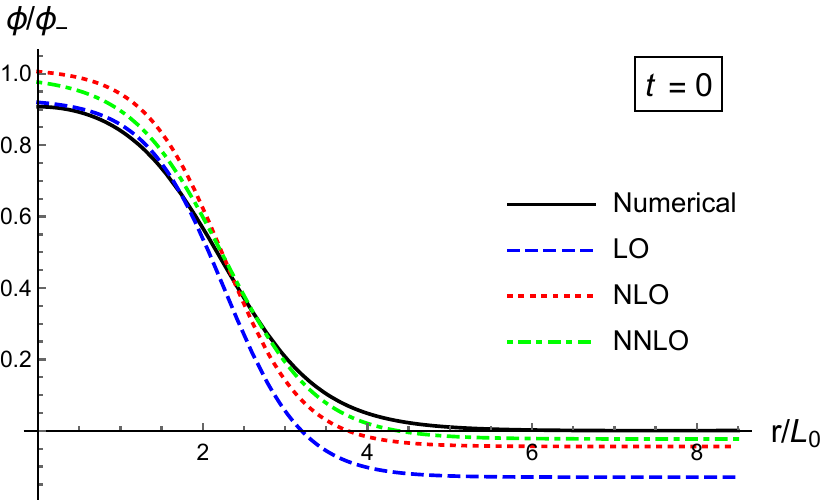}
\hfill
\includegraphics[width=0.49\textwidth]{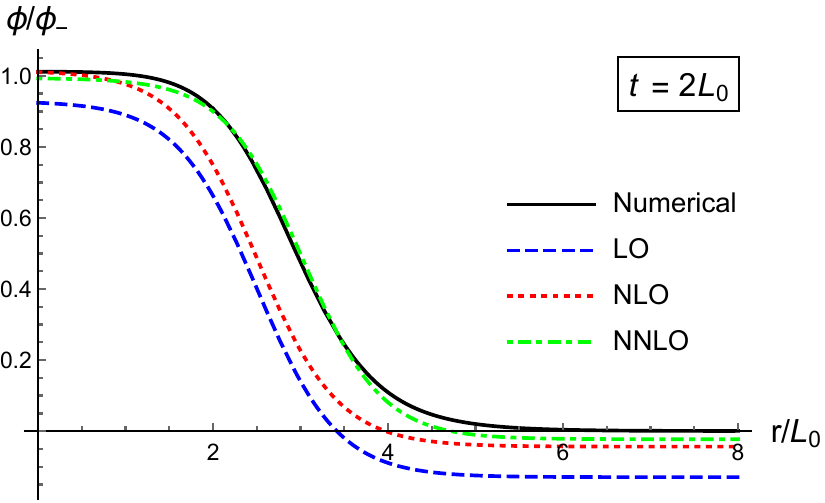}
\hfill
\includegraphics[width=0.49\textwidth]{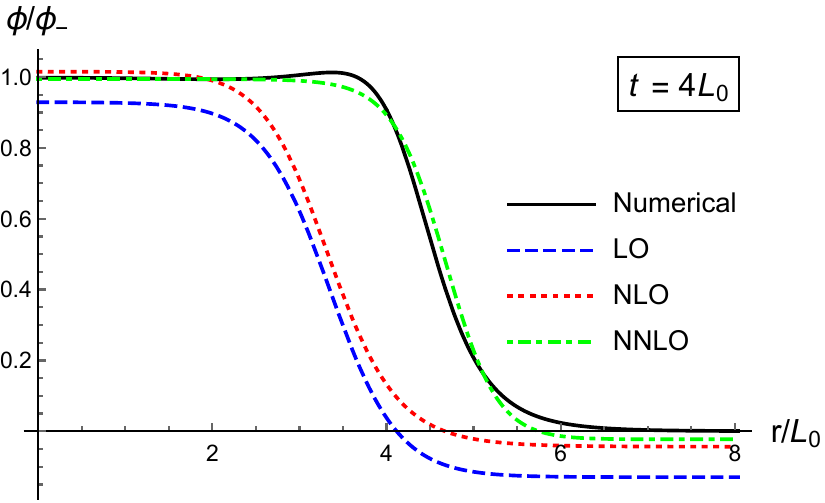}
\hfill
\includegraphics[width=0.49\textwidth]{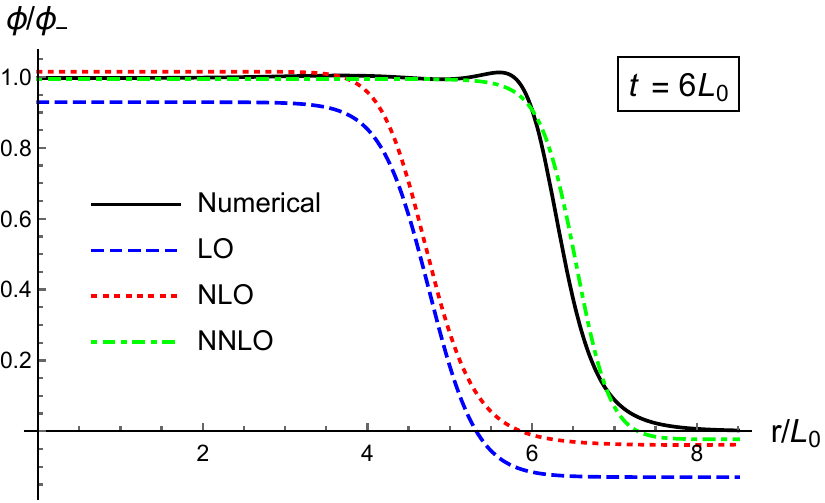}
\caption{The numerical profile and the approximations for the potential shape
shown in the right panel of Fig.~\ref{fig:pot} ($\Delta V/V_{b}\simeq42$,
$l_{0}/L_{0}\simeq0.54$).\label{fig:proftmg}}
\end{figure}

In the right panel of Fig.~\ref{fig:rw} and in Fig.~\ref{fig:proftmg}, we consider the  case
$\Delta V/V_{b}\simeq42$. 
We see that the NNLO approximation for $r_w$ works quite well.
The maximum relative deviation from the numerical value $\bar{r}_{w}$ is about 3\% (at $t\simeq3.8 L_{0}$).
In contrast, the LO (and NLO) approximation quickly departs from the correct value.
Nevertheless, at higher $t$, all the curves in Fig.~\ref{fig:rw} become parallel, so the relative differences decrease with time.
It is interesting to note that, if we use instead the initial conditions $r_{i0},r_i$ estimated with the thin-wall approximation at each order,
the evolution of the approximations is the opposite, i.e., the initial values are different from the numerical value, but the curves approach each other at later times (see Sec.~\ref{sec:nucleation}).

In Fig.~\ref{fig:proftmg} we consider the profile at several times.
As shown by the black curve in the first plot, 
the field does not initially take the value $\phi=\phi_{-}$ inside the nucleated bubble.
In contrast, the thin-wall approximation does approach this value (with increasing precision at higher
order), and thus fails to reproduce the initial moments of the bubble profile
(the fact that the LO curve approaches the value $\phi_-$ is just a numerical coincidence).
At later times, the field inside the bubble is in the potential minimum,
and we see that the NNLO curve gives a much better approximation
than the LO and NLO curves. 
Notice, however, that the main error in these curves is due to the wall position
rather than to the shape of the profile.

\subsection{Perturbations on a planar wall}

\begin{figure}[tb]
\centering
\includegraphics[width=0.49\textwidth]{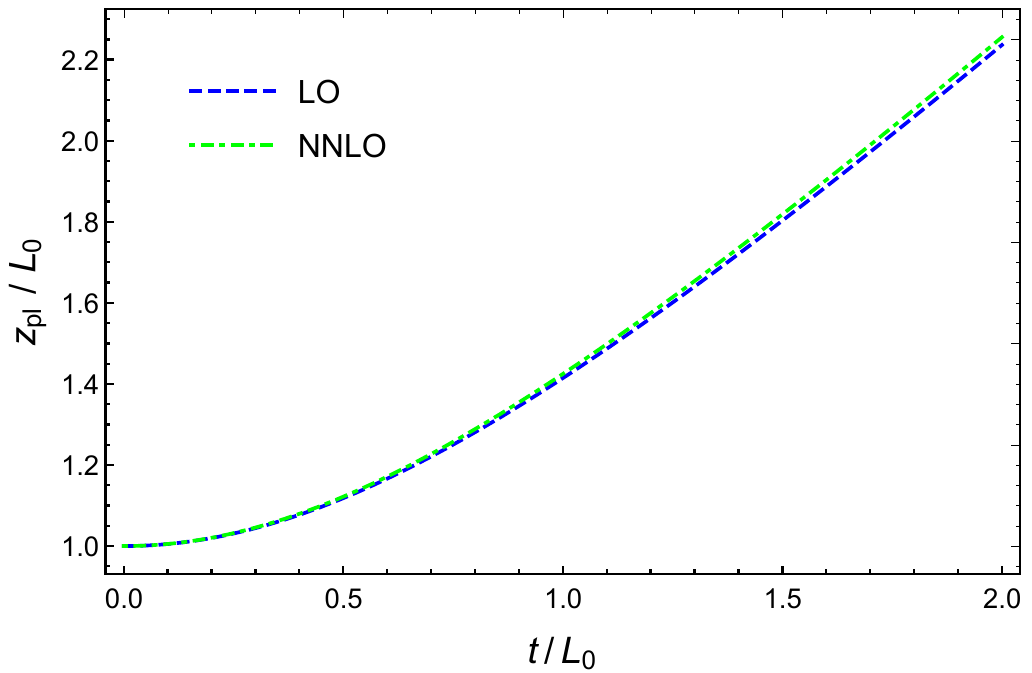}
\hfill
\includegraphics[width=0.49\textwidth]{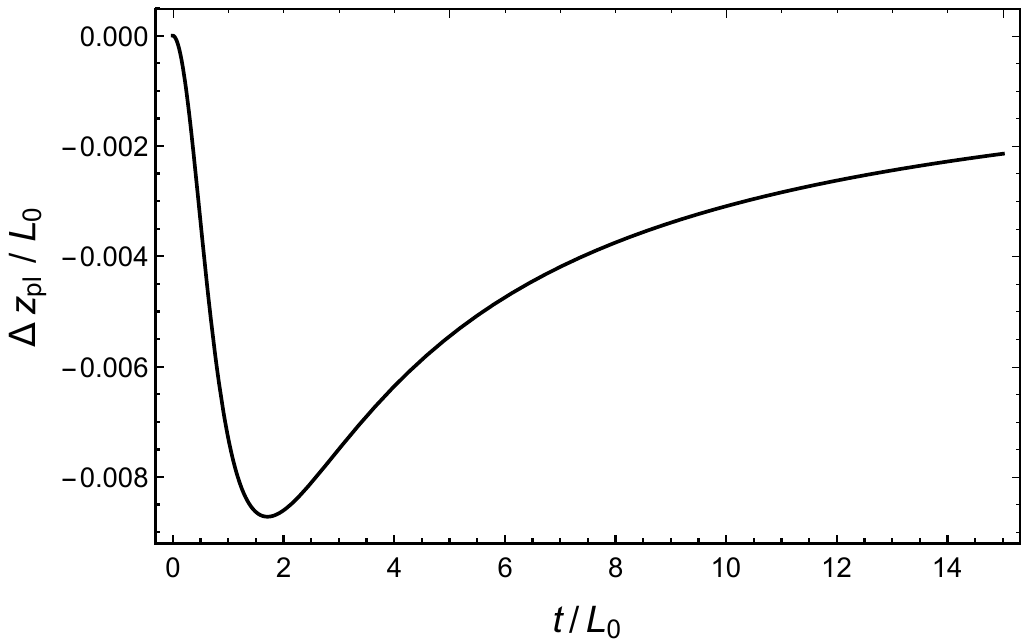}
\caption{LO and NNLO solutions for the position $z_{\mathrm{pl}}$ of a
planar wall (left) and their difference $\Delta z_{\mathrm{pl}}$
(right).\label{fig:planar}}
\end{figure}

Let us now consider small sinusoidal perturbations from a planar wall, for
the concrete case of the potential with $\Delta V/V_{b}\simeq3.5$,
for which we have  $l_{0}/L_{0}\simeq0.3$.
The unperturbed solution for the wall position is given by Eq.~(\ref{eq:solpl0})
at leading order in the wall thickness and by Eq.~(\ref{eq:solpl2})
at next-to-next-to-leading order. 
We set the initial conditions $z_{i}=z_{i0}=L_{0}$, $\dot{z}_{i}=\dot{z}_{i0}=0$.
These background solutions are shown
in Fig.~\ref{fig:planar}. We see that the NNLO correction
to the dynamics is smaller than for the spherical wall (cf.~left
panel of Fig.~\ref{fig:rw}). 

\begin{figure}[tb]
\includegraphics[width=0.49\textwidth]{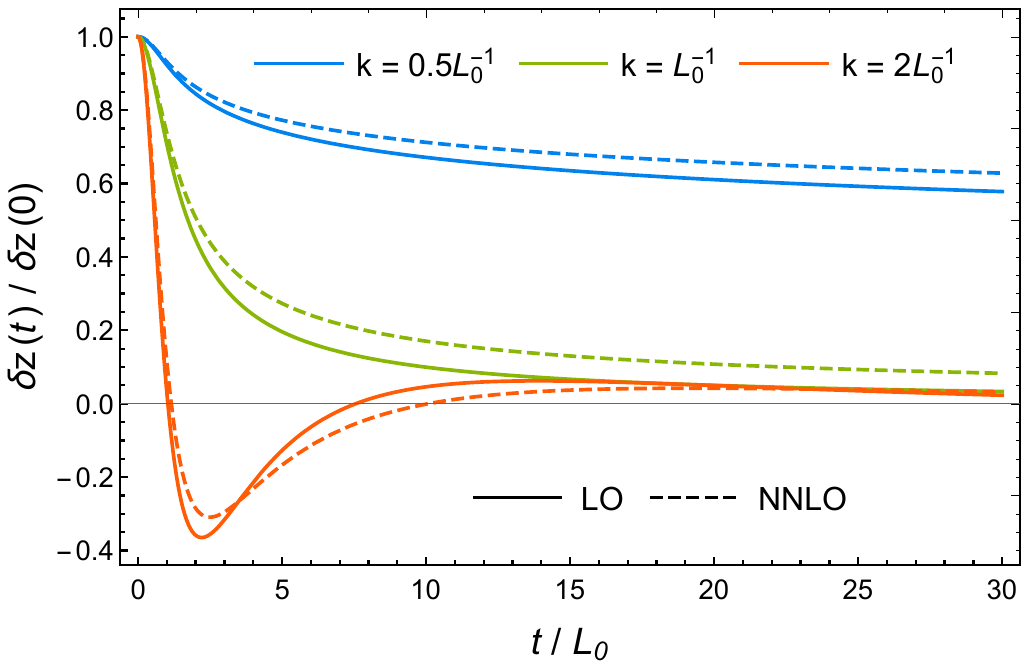}
\hfill
\includegraphics[width=0.49\textwidth]{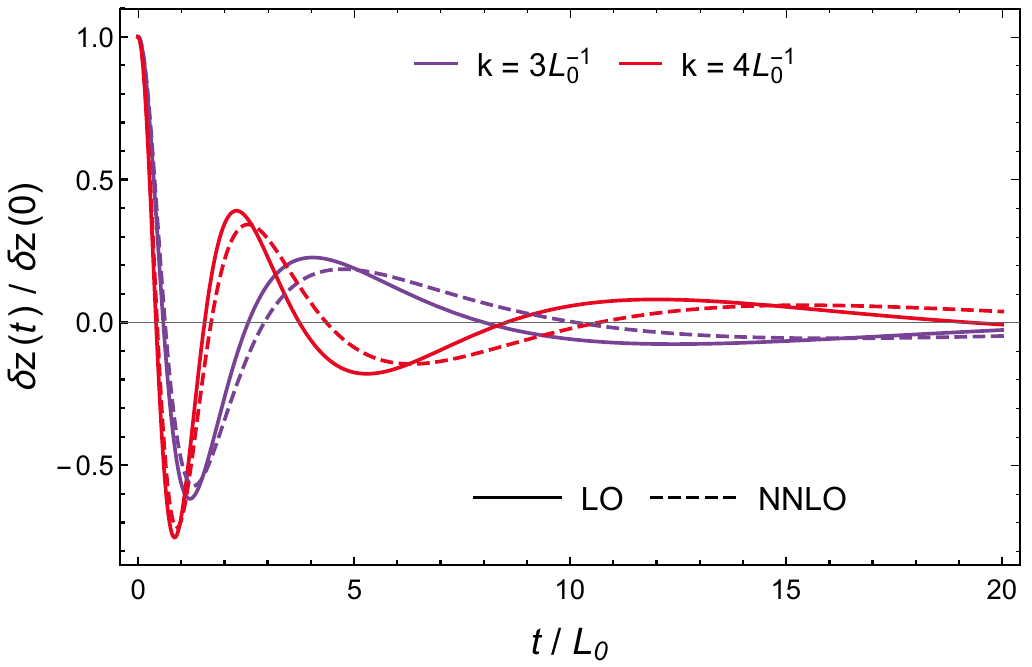}
\caption{The amplitude of sinusoidal deformations.\label{fig:deltaz}}
\end{figure}

For the perturbations, we shall also consider the same initial conditions at each order, namely,
$\delta z=\delta z_{0}$, with $\delta\dot{z}=\delta\dot{z}_{0}=0$.
The LO approximation $\delta z_{0}$ is given by Eq.~(\ref{eq:deltazk0})
and the NNLO approximation $\delta z$ is given by Eq.~(\ref{eq:deltazk2}).
In Fig.~\ref{fig:deltaz}, we show the evolution of the amplitude
of the perturbations for a few values of the wavenumber
$k$. The NNLO correction is quite larger than in the background
solution. This is because the curvature of the hypersurface is no
longer given by the acceleration alone (which would give $K=L_{0}^{-1}$),
but the undulations of the surface introduce a curvature of order
$k^{-1}$. Therefore, the expansion in the wall width is an expansion
in powers of $l_{0}k$ as well as an expansion in $l_{0}/L_{0}$.
For the modes considered in Fig.~\ref{fig:deltaz}, we have $l_{0}k\simeq0.15$,
0.3, 0.6, 0.9 and 1.2.

\begin{figure}[tb]
	\centering
	\includegraphics[width=0.49\textwidth]{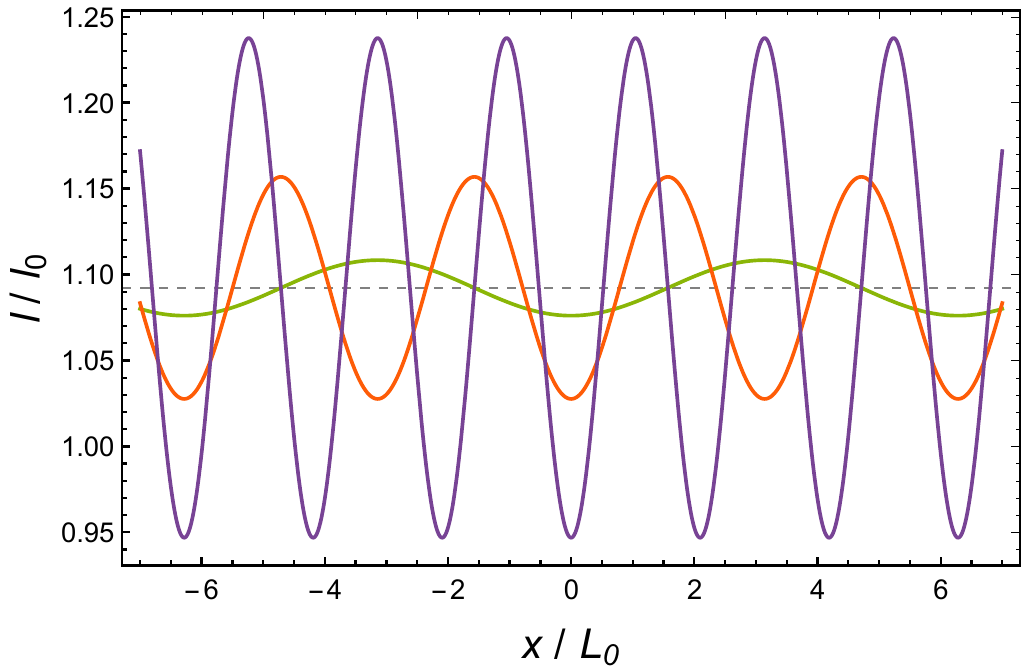}
	\hfill
	\includegraphics[width=0.49\textwidth]{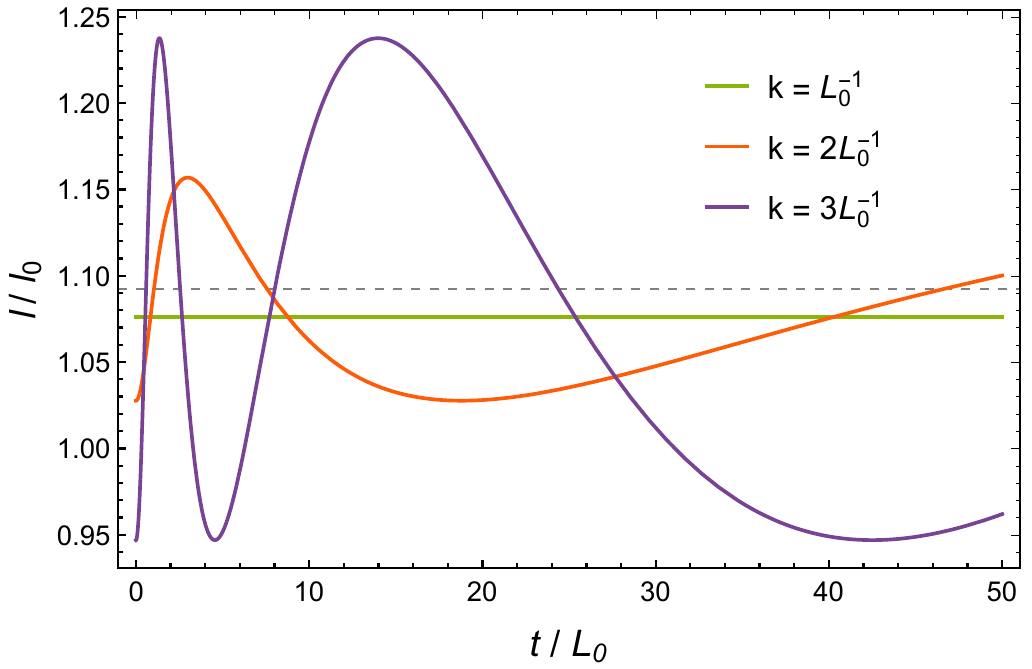}
	\caption{Variations of the wall width $l$ when the surface has sinusoidal
		deformations of amplitude $A_{0}=0.1L_{0}$ ($A_{0}/l_{0}\simeq0.34$).\label{fig:deltal}}
\end{figure}

Using the result (\ref{eq:dnK0k}) in Eq.~(\ref{eq:ltotpot}), we obtain the variations of
$l$,
\begin{equation}
	\frac{l}{l_{0}}=1+\frac{3c^{2}}{8\lambda^{2}a^{6}}
	+\frac{0.590948}{\lambda a^{2}L_{0}^2}
	 \left[1-2L_{0}A_{0}k^{2}\cos\left[\omega_{0}L_{0}\mathrm{arcsinh}(t/L_{0})\right]\cos(\mathbf{k}\cdot\mathbf{x})\right]. 
	 \label{eq:lt}
\end{equation}
Besides the constant corrections, which are due to the global acceleration, 
there are corrections due to the varying curvature
of the hypersurface. As a consequence, $l$ oscillates around a fixed
value $\bar{l}$. Fig.~\ref{fig:deltal} shows the variation of $l$ along the wall surface at $t=0$ and its time variation
at a given point (the value $\bar{l}$ is indicated by a dashed line). 
We see that, the larger the curvature, the stronger the effect.
This is to be expected since the variations of the
wall width are a consequence of the surface bending. 

\section{The O(3,1)-invariant bubble}

\label{sec:nucleation}

The process of bubble nucleation is described by the bounce instanton \cite{c77}.
The most probable configuration of the bubble at the moment of its materialization
corresponds to the bounce of minimum Euclidean action.
The latter is O(4) invariant, and the initial evolution of the bubble, which is the analytic continuation of the bounce,
is O(3,1) invariant.
Hence, the field is of the form $\phi(x^{\mu})=\bar{\phi}(\rho)$,
with $\rho=\sqrt{r^{2}-t^{2}}$. In Ref.~\cite{c77}, Coleman considered
the leading-order thin-wall approximation for the instanton and applied
it to $\bar{\phi}(\rho)$. Here, we shall consider higher orders in
the wall width. 
While in Sec.~\ref{sec:pert_method} we obtained
a wall EOM which required initial conditions,
imposing a symmetric solution will determine the initial bubble radius
and wall velocity.

In Sec.~\ref{sec:pert_method}, we used normal Gaussian
coordinates because the field depends only on the variable $n$ at the lowest orders in the wall width.
However, if we assume that the exact solution depends only on $\rho$, the dependence on a single variable
must be valid at every order in the wall width. 
In particular, the exact
field equation in terms of $\rho$, which is given by Eq.~(\ref{eq:ecColeman}),
takes the same form of the approximation (\ref{eq:primerasup}),
with $K=-3/\rho$. 
We need to adapt our method to use the variable $\rho$ instead of $n$.
This will give an explicitly Lorentz-invariant solution at each order. 

With the assumption that $\phi$ is a function of  $\rho$ only, the treatment is much simpler than
in Sec.~\ref{sec:pert_method}. 
A fixed point in the field profile
corresponds to a fixed value of $\rho$. 
Therefore, if we define the wall hypersurface by $\phi(x^{\mu})=\phi_{w}$,
where $\phi_{w}$ is some reference value between $\phi_{+}$ and
$\phi_{-}$,
we have the condition $\rho(x^{\mu})=\rho_{w}$,
where $\bar{\phi}(\rho_{w})=\phi_{w}$. 
This condition determines the
form of the solution for the bubble radius, namely, $r=r_{w}(t)$,
with
\begin{equation}
	r_{w}^{2}=\rho_{w}^{2}+t^{2}.\label{eq:solColeman}
\end{equation}
Thus, we do not need to solve an equation of motion but only calculate
the value of the constant $\rho_{w}$, which also gives the initial
bubble radius.

We may define the reference value $\rho_{w}$ as the weighted average
\begin{equation}
	\rho_{w}=\sigma^{-1}\int_{0}^{\infty}\bar{\phi}'(\rho)^{2}\rho\,d\rho,\label{eq:defrhoc}
\end{equation}
where $\sigma=\int_{0}^{\infty}\bar{\phi}'(\rho)^{2}d\rho$. 
As usual,
we will make the approximation that the field takes the value $\phi=\phi_{-}$ inside the bubble. 
This value will be reached asymptotically
at $\rho=-\infty$ instead of $\rho=0$. 
Therefore, we must replace the limit
of integration $\rho=0$ with $\rho=-\infty$ in the expressions above.
For a discussion of these approximations, see \cite{mm23}. The role
of the variable $n$ in Sec.~\ref{sec:pert_method} is played here
by $\eta=\rho-\rho_{w}$. Indeed, in terms of $\eta$, Eq.~(\ref{eq:defrhoc})
becomes $\int_{-\infty}^{+\infty}\bar{\phi}^{\prime2}\eta d\eta=0$,
which is similar to the condition (\ref{eq:cero_n}).

Multiplying Eq.~(\ref{eq:ecColeman}) by $d\bar{\phi}/d\rho$ and
integrating, we obtain a first integral which, in terms of $\eta$,
takes the form
\begin{equation}
	\frac{1}{2}\left(\frac{d\bar{\phi}}{d\eta}\right)^{2}
	+\int_{\infty}^{\eta}\frac{3}{\rho_{w}+\tilde{\eta}}\left(\frac{d\bar{\phi}}{d\tilde{\eta}}\right)^{2}d\tilde{\eta}
	=V(\bar{\phi})-V_{+}.\label{eq:primeraintColeman}
\end{equation}
Evaluating Eq.~(\ref{eq:primeraintColeman}) at $\eta=-\infty$,
we obtain the equation 
\begin{equation}
	\int_{-\infty}^{+\infty}\frac{3}{\rho_{w}+\eta}\left(\frac{d\bar{\phi}}{d\eta}\right)^{2}d\eta=\Delta V.\label{eq:eomColeman}
\end{equation}
The last two equations replace Eqs.~(\ref{eq:primeraint}) and (\ref{eq:eom}).
To lowest order in the wall width, we neglect the term proportional
to $K=-3/\rho$ in Eq.~(\ref{eq:primeraintColeman}) and replace
$V$ with the degenerate potential $V_{0}$. Thus, Eq.~(\ref{eq:primeraintColeman})
takes the form of Eq.~(\ref{eq:primeraint0}), and we obtain the
solution $\bar{\phi}(\rho)=\phi_{0}(\eta)$, with $\phi_{0}$ given
by Eq.~(\ref{eq:fi0}). Like in Sec.~\ref{sec:pert_method}, to
calculate $\bar{\phi}$ to higher order, we write
\begin{equation}
	\bar{\phi}(\rho)=\phi_{0}(\eta)+\phi_{1}(\eta)+\phi_{2}(\eta)+\cdots.
\end{equation}
We
also expand of $3/\rho$ as a series in $\eta$ and separate $\rho_{w}$ into terms of
different orders,
\begin{equation}
	\frac{3}{\rho_{w}+\eta}=\frac{3}{\rho_{w}}-\frac{3}{\rho_{w}^{2}}\eta+\cdots,\qquad\rho_{w}=\rho_{0}+\rho_{1}+\cdots.
\end{equation}

Inserting these expansions in Eq.~(\ref{eq:primeraintColeman}),
we obtain Eqs.~(\ref{eq:primeraintpert}) for $\phi_{i}(\eta)$,
with the quantities $f_{i}$ given by Eqs.~(\ref{f1})-(\ref{f3}),
but with 
\begin{align}
	K_{0}|_{0} & =-\frac{3}{\rho_{0}}, \ K_{1}|_{0}=\frac{3\rho_{1}}{\rho_{0}^{2}}, \  K_{2}|_{0}=\frac{3\rho_{2}}{\rho_{0}^{2}}-\frac{3\rho_{1}^{2}}{\rho_{0}^{3}}, \\
	\partial_{n}K_{0}|_{0} & =\frac{3}{\rho_{0}^{2}}, \ \partial_{n}K_{1}|_{0}=-\frac{6\rho_{1}}{\rho_{0}^{3}}, \ \partial_{n}^{2}K_{0}|_{0}=-\frac{6}{\rho_{0}^{3}}.
	\label{eq:K_Coleman}
\end{align}
The functions $\phi_{i}(\eta)$ are thus given by the solution (\ref{eq:solfii}), but there is no dependence on $\xi^{a}$. On
the other hand, evaluating Eq.~(\ref{eq:primeraintpert}) at $\eta=-\infty$ gives $f_i=0$,
which is equivalent to considering the perturbative expansion of Eq.~(\ref{eq:eomColeman}).
We obtain 
\begin{equation}
	\frac{3\sigma_{0}}{\rho_{0}}=\Delta V_{1}, \quad
	\frac{3\sigma_{1}}{\rho_{0}}-\frac{3\rho_{1}}{\rho_{0}^{2}}\sigma_{0}=\Delta V_{2}, \quad
	\frac{3\sigma_{2}}{\rho_{0}}-\frac{3\rho_{1}\sigma_{1}}{\rho_{0}^{2}}-\frac{3\rho_{2}\sigma_{0}}{\rho_{0}^{2}}+\frac{3\rho_{1}^{2}\sigma_{0}}{\rho_{0}^{3}}+\frac{3\mu_{0}}{\rho_{0}^{3}}=\Delta V_{3},\label{eq:eomi_coleman}
\end{equation}
with $\Delta V_{i}$ defined below Eqs.~(\ref{V1})-(\ref{V3})
and $\sigma_{i},\mu_{i}$ defined below Eq.~(\ref{eq:Gi}). The first
of Eqs.~(\ref{eq:eomi_coleman}) gives the radius $\rho_{w}$ to
leading order, $\rho_{0}=3\sigma_{0}/\Delta V_{1}$. 
Hence, to this order, we can write $\rho_{w}=3\sigma_{0}/\Delta V=R_{0}$, which is the
well known result. Adding the first two equations in (\ref{eq:eomi_coleman}), we obtain
\begin{equation}
	\rho_{0}+\rho_{1}\simeq\frac{3\left(\sigma_{0}+\sigma_{1}\right)}{\Delta V_{1}+\Delta V_{2}},
\end{equation}
which again gives $\rho_{w}=3\sigma/\Delta V$. Adding the three equations,
we obtain 
\begin{equation}
	\rho_{w}\simeq\frac{3\left(\sigma_{0}+\sigma_{1}+\sigma_{2}\right)}{\Delta V_{1}+\Delta V_{2}+\Delta V_{3}}+\frac{\Delta V_{1}}{3\sigma_{0}}\frac{\mu_{0}}{\sigma_{0}}\simeq\frac{3\sigma}{\Delta V}+\frac{l_{0}^{2}}{\rho_{0}}.
\end{equation}
The quantities $\sigma_{i}$ and $\mu_{i}$ have the same expressions as in Sec.~\ref{sec:pert_method}.
In particular, we have $\sigma_{2}=\tilde{\sigma}_{2}-\mu_{0}\partial_{n}K_{0}$,
where, in this case, we have $\partial_{n}K_{0}|_{0}=3/\rho_{0}^{2}$.
Hence, we have
\begin{equation}
	\rho_{w}=\frac{3\tilde{\sigma}}{\Delta V}-2\frac{l_{0}^{2}}{\rho_{0}}\equiv R.\label{eq:rhoc}
\end{equation}
Finally, the field profile is given by Eq.~(\ref{eq:fitot}), where the functions $\phi_{i}$ are the same as in Sec.~\ref{sec:pert_method} with $n$ replaced by $\eta$, and $\partial_{n}K_{0}|_{0}=3/\rho_{0}^{2}$ .

\begin{figure}[tb]
	\centering
	\includegraphics[width=0.45\textwidth]{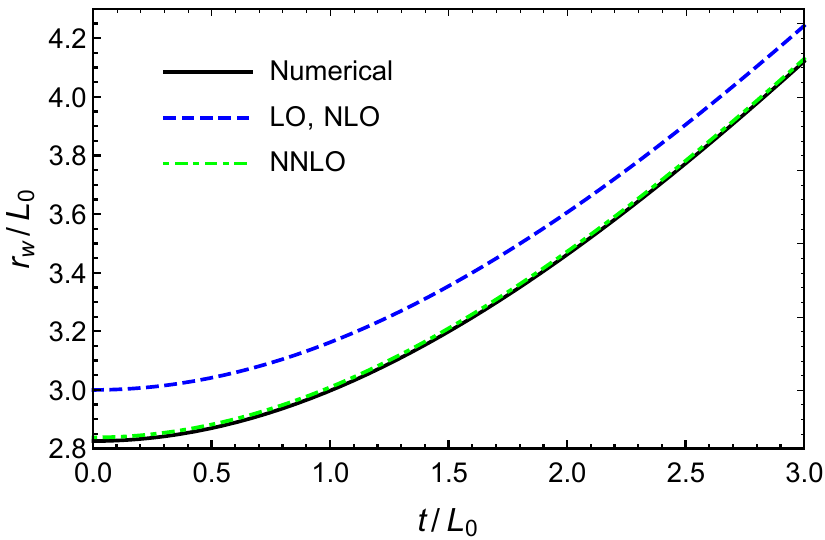}
	\hfill
	\includegraphics[width=0.45\textwidth]{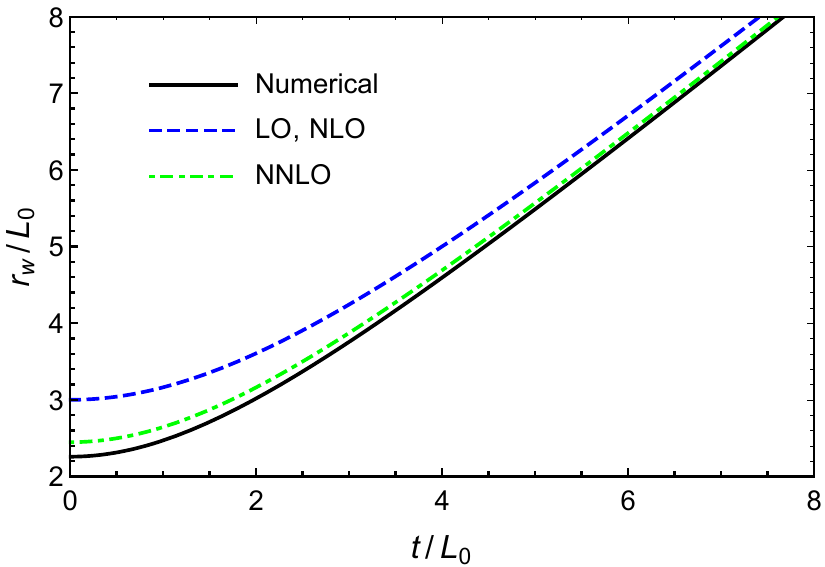}
	\caption{The bubble radius and its approximations for the two potential shapes
		shown in Fig.~\ref{fig:pot} (cf.~Fig.~\ref{fig:rw}).\label{fig:rwcol}}
\end{figure}

As already discussed, the O(3,1) solution is the analytic continuation
of the O(4) bounce instanton. 
We could have equivalently considered the Euclidean case by using the imaginary time version of $\rho$, namely, 
$\rho=\sqrt{\tau^2+r^2}$. Thus, $\rho_{w}$ gives the radius of
the nucleated bubble. 
For the potential considered in Sec.~\ref{sec:ejemplos},  Eq.~(\ref{eq:defrhoc}) gives a numerical value 
$\rho_w \simeq 2.83 L_0$ for the case with $\Delta V/V_{b}\simeq3.5$ and $\rho_w\simeq 2.26L_0$ for the case with $\Delta V/V_{b}\simeq42$.
The leading order approximation $\rho_{w}=R_{0}=3L_0$ has an error of 6\% in the first case and of 33\% in the second case,
while the NNLO approximation $\rho_{w}=R$
has an error of 0.5\% in the first case and of 8\% in the second case%
\footnote{On the other hand, the approximation $\rho_{w}=\tilde{R}$ and the
	analytic approximation for $r_{w}(t)$ given in footnote \ref{fn:NNLOanalit}
	have an error of approximately 2.5\% for the case in the left panel
	of Fig.~\ref{fig:rwcol} and 17\% for the case in the right panel.
	The corresponding curves are halfway between the LO and NNLO results
	of Fig.~\ref{fig:rwcol}.}.
In Fig.~\ref{fig:rwcol} we plot the bubble radius $r_{w}=\sqrt{\rho_{w}^{2}+t^{2}}$ 
for the different approximations for $\rho_w$.
Notice that the radius $\bar{r}_{w}(t)$, defined in Eq.~(\ref{eq:rmedio})
as a spatial average, will not coincide exactly with this Lorentz-invariant form.
It corresponds to a different definition of the wall position. 
Nevertheless, we have checked that the black curves in Fig.~\ref{fig:rw}
are almost coincident with the ones in Fig.~\ref{fig:rwcol}. 

\section{Conclusions}

\label{sec:conclu}

This work is the second of a series in which we aim to derive the
equation of motion of a bubble wall, as well as the wall profile,
under the most general conditions. In Ref.~\cite{mm23}, we considered
an infinitely thin wall of arbitrary shape in a vacuum phase transition,
and in a forthcoming paper we will consider a thermal phase transition.
However, the usual thin-wall approximation breaks down if the worldvolume
bends in such a way that the curvature radius becomes of the order
of the wall width (which may happen, in particular, for an accelerated
planar or spherical wall). In this paper, we have presented a method
to obtain the wall profile and the wall EOM perturbatively in the wall width.
For simplicity, we have considered a vacuum phase transition, but
the generalization is straightforward.

The usual thin-wall approximation assumes a bubble profile of the
form $\phi=\phi(n)$, where $n$ is the distance perpendicular to
the wall hypersurface. In a coordinate system where this variable
is a coordinate, such a rigid profile has a fixed width $l$ and a
constant surface tension $\sigma$. Assuming also that quantities
such as the metric vary little across the thin wall,
one obtains an EOM for the wall surface which
depends on the ratio $\Delta V/\sigma$.
Assuming that the potential is almost degenerate, 
one also obtains an estimation
for the parameter $\sigma$. In Ref.~\cite{mm23}, we showed that all
the above remains valid at next-to-leading order in the wall width,
and we calculated the corrections to the LO profile $\phi_{0}(n)$
and surface tension $\sigma_{0}$. In the present paper, we have derived
a systematic calculation of the field profile and the wall EOM at
higher orders in the wall width. As we have seen, at the next-to-next-to-leading
order, $\phi$ no longer depends solely on $n$. Thus, the corrections
to the quantity $\sigma$ are no longer constant. Besides, new parameters
appear in the wall EOM, such as the wall width $l$. This quantity,
which we concretely define as the root mean square value of $n$ weighted
with the function $\left(\partial_{n}\phi\right)^{2}$, also varies
along the hypersurface beyond the next-to-leading order.

To summarize our method, we first write the field equation in normal
Gaussian coordinates $n,\xi^{a}$. Thus, the zeroth-thickness approximation
corresponds to 1) neglecting the dependence of $\phi$ on $\xi^{a}$,
2) neglecting the mean curvature of the hypersurface $K|_0$ as well as the
variation of the quantity $K$ with $n$, and 3) approximating the potential $V$
by a degenerate potential $V_{0}$. Then,
we essentially consider the equation for $\phi$ as an expansion in
$K$, $V-V_{0}$, and the
successive derivatives $\partial_{a}\phi$, $\partial_{n}K$. This
procedure gives simple equations for $\phi$ at each order, which
can be easily solved. All these equations have the same form beyond
the leading order, and we found their general solution. Each correction
to $\phi$ is given by a combination of terms of the form $f(n)\partial_{n}^{k}K(\xi^{a})$
and can be readily put in a covariant form. On the other hand, integrating
the field equation with respect to $n$ gives the wall EOM at each
order. The LO and NLO equations are of the form $\sigma K=\Delta V$,
while at higher orders there are also terms with derivatives $\partial_{n}^{k}K$.

Small modifications of this treatment can be convenient for certain cases.
As an example, we have considered the O(3,1)-symmetric bubble or the corresponding O(4)-symmetric instanton.
The symmetry of the problem makes it convenient to
replace the coordinates $n,\xi^a$ with a single variable $\rho=\sqrt{r^2-t^2}$. 
This approach is a direct generalization of Coleman's treatment \cite{c77} and 
gives the finite-width corrections to the nucleation radius.

For a simple case such as that of a spherical bubble, 
a numerical computation of the field profile is straightforward, 
and we have considered this case to test our results.
As we have seen, the NNLO solution  improves significantly
the NLO one and gives an excellent approximation
both for the profile and the wall position, even for potentials which depart
significantly from the degenerate approximation.
For a general wall evolution, 
a numerical approach such as a lattice calculation can be quite computationally intensive.
On the other hand, our analytic expansion can be cumbersome, but the numerical evaluation of the expressions will generally be straightforward.
In particular, for a polynomial potential, we have obtained analytic
expressions for the field profile and the coefficients of the EOM.
For a planar wall with small deformations, we have obtained analytic solutions for the evolution of the perturbations beyond the zero-thickness approximation.

\section*{Acknowledgements}

This work was supported by Universidad Nacional de Mar del Plata,
grant EXA1091/22.

\appendix

\section{Normal Gaussian coordinates}

\label{sec:gaussian}

In this appendix we review the construction and properties
of normal Gaussian coordinates and the extrinsic curvature tensor
of a hypersurface (for details, see, e.g., \cite{synge,wald,carroll}).

Given a point $X^{\mu}(\xi^{a})$ on the hypersurface $\Sigma$ and the normal vector
$N^{\mu}$, let us consider a geodesic that crosses $\Sigma$ perpendicularly.
The general equation for a geodesic is
\begin{equation}
	\frac{d^{2}x^{\mu}}{dn^{2}}+\Gamma_{\nu\rho}^{\mu}\frac{dx^{\nu}}{dn}\frac{dx^{\rho}}{dn}=0,\label{eq:geod}
\end{equation}
where $n$ is an arbitrary parameter and
$\Gamma_{\alpha\beta}^{\mu}=\frac{1}{2}g^{\mu\nu}
(\frac{\partial g_{\nu\alpha}}{\partial x^{\beta}}+\frac{\partial g_{\nu\beta}}{\partial x^{\alpha}}-\frac{\partial g_{\alpha\beta}}{\partial x^{\nu}})$ 
are the Christoffel symbols.
In our case, we have the initial conditions
\begin{equation}
x^{\mu}(0)=X^{\mu}(\xi^{a}),\quad\frac{dx^{\mu}}{dn}(0)=N^{\mu}(\xi^{a}),\label{eq:cc_geod}
\end{equation}
and $n$ becomes the proper distance. 
Indeed, Eq.~(\ref{eq:geod}) has the first integral
\begin{equation}
g_{\mu\nu}\frac{dx^{\nu}}{dn}\frac{dx^{\mu}}{dn}=-1,\label{eq:normgeod}
\end{equation}
where the specific value of the constant comes from the second condition (\ref{eq:cc_geod})
and the normalization of $N^{\mu}$. As a consequence, the line element
on the geodesic is given by $ds^{2}=g_{\mu\nu}dx^{\mu}dx^{\nu}=-dn^{2}$. 
According to Eq.~(\ref{eq:normgeod}), in the coordinates $\bar{x}^{\mu}=(\xi^{a},n)$
we have $\bar{g}_{nn}=-1$ along the geodesic. Near $\Sigma$, the
change of coordinates is given by Eq.~(\ref{eq:Gaussian}), which
implies that $\bar{g}_{an}|_{n=0}=g_{\mu\nu}\partial_{a}X^{\nu}N^{\mu}=0$.
Moreover, $\bar{g}_{an}$ is constant along the geodesic. This can
be seen from the fact that, in normal Gaussian coordinates, Eq.~(\ref{eq:geod})
takes the form $\bar{\Gamma}_{nn}^{\mu}=0$, which implies $\partial_{n}\bar{g}_{an}=0$.
Hence, we have $\bar{g}_{an}=0$ also away from $\Sigma$. The inverse
matrix $\bar{g}^{\mu\nu}$ also fulfills $\bar{g}^{nn}=-1$, $\bar{g}^{an}=0$,
and we have $\bar{g}^{ac}\bar{g}_{cb}=\delta_{b}^{a}$. It is 
easy to verify that the Christoffel symbols are given by Eq.~(\ref{eq:GammaGauss}).

We define the vector field 
\begin{equation}
n^{\mu}= dx^{\mu}/dn,
\end{equation}
which is tangent to the geodesics and coincides with $N^{\mu}$
on $\Sigma$. According to Eq.~(\ref{eq:normgeod}), it satisfies
$n_{\mu}n^{\mu}=-1$ everywhere. Furthermore, the property $\bar{g}_{an}=0$
implies that we have $\partial_{a}x^{\mu}n_{\mu}=0$ everywhere. These
properties indicate that $n^{\mu}$ is a normal vector to the surfaces $\Sigma_{n}$
of $n=$ constant. In normal Gaussian
coordinates we have 
\begin{equation}
\bar{n}^{\mu}=(0,0,0,1),\quad\bar{n}_{\mu}=(0,0,0,-1).\label{eq:nmugauss}
\end{equation}
We may write the latter covariantly as ${n}_{\mu}=-\partial_\mu n$.
The vector $N_{\mu}=-\partial_{\mu}F/s$ coincides with $n_{\mu}$ only
at $n=0$, i.e., on $\Sigma$. 
Since both vectors are normalized,
we have $N^{\nu}\nabla_{\mu}N_{\nu}=0$ and $n^{\nu}\nabla_{\mu}n_{\nu}=0$.
On the other hand, $n^{\mu}$ also satisfies $n^{\mu}\nabla_{\mu}n_{\nu}=0$,
as can be easily seen from Eqs.~(\ref{eq:nmugauss}) and (\ref{eq:GammaGauss}).

To obtain the expression for  $n$ as a function of $x^{\mu}$, we
begin by considering the expansion 
$F=an+bn^{2}+cn^{3}+\cdots$,
where $a=\partial_{n}F|_{n=0}$, $b=\frac{1}{2}\partial_{n}^{2}F|_{n=0}$,
and so on. 
The inverse of this expansion is 
\begin{equation}
	n=F/a-(b/a)\left(F/a\right)^{2}+\left[2\left(b/a\right)^{2}-c/a\right]\left(F/a\right)^{3}+\cdots.\label{eq:nexpanF}
\end{equation}
Taking into account that $\partial_{n}F=n^{\mu}\partial_{\mu}F=-sn^{\mu}N_{\mu}$
and similar equations for successive derivatives, then evaluating
at $n=0$, we obtain
\begin{equation}
a=s|_{n=0},\quad b=\frac{1}{2}N^{\nu}\partial_{\nu}s|_{n=0},\quad c=\frac{1}{6}\left[N^{\rho}N^{\nu}\nabla_{\rho}\partial_{\nu}s-sN^{\rho}N^{\mu}N^{\nu}\nabla_{\rho}\nabla_{\nu}N_{\mu}\right]_{n=0}.\label{abc}
\end{equation}
Inserting  the coefficients (\ref{abc}) in Eq.~(\ref{eq:nexpanF}), we obtain
\begin{equation}
n=c_{1}(\xi^{a},0)F+c_{2}(\xi^{a},0)F^{2}+c_{3}(\xi^{a},0)F^{3}+\cdots,\label{eq:nsegundoorden}
\end{equation}
with
\begin{equation}
c_{1}=\frac{1}{s},\quad c_{2}=-\frac{N^{\nu}\partial_{\nu}s}{2s^{3}},\quad c_{3}=\frac{\left(N^{\nu}\partial_{\nu}s\right)^{2}}{2s^{5}}+\frac{N^{\rho}N^{\nu}N^{\mu}\nabla_{\rho}\nabla_{\nu}\left(sN_{\mu}\right)}{6s^{4}}.\label{eq:ci}
\end{equation}
The expression (\ref{eq:nsegundoorden}) gives an expansion with coefficients
evaluated at the hypersurface $\Sigma$. To obtain $n$ as a function
of the coordinates $x^{\mu}$, we may expand these coefficients as
\begin{equation}
c_{i}(\xi^{a},0)=c_{i}(\xi^{a},n)-\partial_{n}c_{i}(\xi^{a},n)n+\frac{1}{2}\partial_{n}^{2}c_{i}(\xi^{a},n)n^{2}+\mathcal{O}(n^{2}).
\label{eq:coefci}
\end{equation}
Using $\partial_{n}c_{i}=N^{\alpha}\partial_{\alpha}c_{i}$, inserting Eq.~(\ref{eq:coefci})
in Eq.~(\ref{eq:nsegundoorden}), and using the latter recursively,
we obtain Eq.~(\ref{eq:n_orden2}). Notice that, although Eq.~(\ref{eq:n_orden2})
looks very similar to the expansion (\ref{eq:nsegundoorden})-(\ref{eq:ci}),
there is a sign difference in the quadratic term. Besides, all the
quantities in Eq.~(\ref{eq:n_orden2}) are evaluated at the
point $x^{\mu}$ and can be obtained from $F(x^{\mu})$ using $s^{2}=-F_{,\alpha}F^{,\alpha}$,
$sN_{\mu}=-\partial_{\mu}F$, and $\partial_{\nu}s=N^{\mu}\nabla_{\nu}\partial_{\mu}F$.

We define the extrinsic curvature tensor of the hypersurfaces
$\Sigma_{n}$ as \cite{wald}
\begin{equation}
K_{\mu\nu}=-\nabla_{\mu}n_{\nu}.\label{Kmunudef}
\end{equation}
From Eq.~(\ref{eq:nmugauss}), we readily see that, in normal Gaussian
coordinates, we have
\begin{equation}
	\bar{K}_{n\mu}=0,\quad\bar{K}_{ab}=-\bar{\Gamma}_{ab}^{n}\label{eq:Kab}
\end{equation}
and, hence, this tensor has the properties
$n^{\nu}K_{\mu\nu}=0$, $K_{\mu\nu}=K_{\nu\mu}$.
We are interested in the mean curvature $K=\bar{g}^{ab}\bar{K}_{ab}=g^{\mu\nu}K_{\mu\nu}$
and its derivatives with respect to $n$. Applying the general equation
$\nabla_{\rho}\nabla_{\nu}A_{\mu}-\nabla_{\nu}\nabla_{\rho}A_{\mu}={R^{\sigma}}_{\mu\rho\nu}A_{\sigma}=0$
(in flat space) to $A_{\mu}=n_{\mu}$, we obtain the equality $\nabla_{\rho}K_{\mu\nu}=\nabla_{\mu}K_{\rho\nu}$.
Using this property in the covariant derivative of $n^{\mu}K_{\mu\nu}=0$,
we obtain 
\begin{equation}
n^{\mu}\nabla_{\mu}K_{\rho\nu}={K_{\rho}}^{\mu}K_{\mu\nu}.\label{dK}
\end{equation}
Taking into account that $n^{\nu}\nabla_{\nu}K=\partial_{n}K$ and
repeating the manipulations for the second derivative, we obtain the
equations
\begin{equation}
\partial_{n}K=K^{\mu\nu}K_{\mu\nu},\quad\partial_{n}^{2}K=2{K^{\mu}}_{\nu}{K^{\nu}}_{\rho}{K^{\rho}}_{\mu}.\label{eq:dnK}
\end{equation}
At $n=0$, we can write these expressions in terms of $N_{\mu}$.
For that aim, we notice that, due to the property $n^{\mu}K_{\mu\nu}=n^{\mu}K_{\nu\mu}=0$,
we can use the projector ${P_\mu}^\nu=\delta_{\mu}^{\nu}+n_{\mu}n^{\nu}$
to write Eq.~(\ref{Kmunudef}) in the form
$K_{\mu\nu}=-{P_\mu}^\rho{P_\nu}^\sigma\nabla_{\rho}n_{\sigma}$.
Due to the tangential projections, it doesn't matter how the vector
$n^{\mu}$ is extended out of $\Sigma$, and we can replace it with
$N^{\mu}$. Hence we obtain $K_{\mu\nu}=-{h_{\mu}}^{\rho}{h_{\nu}}^{\sigma}\nabla_{\rho}N_{\sigma}$,
with ${h_{\mu}}^{\nu}=\left(\delta_{\mu}^{\nu}+N_{\mu}N^{\nu}\right)$.
Furthermore, since our vector $N_{\mu}$ keeps the normalization $N_{\mu}N^{\mu}=-1$
even out of $\Sigma$, we have $N^{\sigma}\nabla_{\rho}N_{\sigma}=0$
and we obtain Eq.~(\ref{Kmunu}). Inserting the latter in Eqs.~(\ref{eq:K})
and (\ref{eq:dnK}), we obtain Eqs.~(\ref{eq:Kcovar})%
\footnote{It is worth remarking that, although $K_{\mu\nu}=-\nabla_\mu n_\nu$
is symmetric, $\nabla_{\mu}N_{\nu}$ is not, and
we have $K_{\mu\nu}=-\nabla_{\mu}N_{\nu} - N_{\mu}N^{\alpha}\nabla_{\alpha}N_{\nu}$. 
Thus, using either $K^{\mu\nu}K_{\mu\nu}$ or $K^{\mu\nu}K_{\nu\mu}$,
we obtain two different expressions for $\partial_{n}K$, namely, $\partial_{n}K=\nabla^{\mu}N^{\nu}\nabla_{\mu}N_{\nu}+N^{\alpha}N^{\beta}\nabla_{\beta}N^{\nu}\nabla_{\alpha}N_{\nu}$
or $\partial_{n}K=\nabla^{\mu}N^{\nu}\nabla_{\nu}N_{\mu}$.}.

\section{Wall profile and EOM parameters}

\label{sec:perfil}

The calculation of the NNLO field profile, and, hence, of parameters such as
the surface tension, involves several functions of $n$ and integrals with respect to this variable. 
We shall now see that these equations become simpler if we use as a variable the LO field $\phi_{0}$.
This can be accomplished through the function $n(\phi_0)$ defined by Eqs.~(\ref{eq:fi0})-(\ref{eq:nast}).
For instance, it is well known that the LO surface tension can be written as
\begin{equation}
	\sigma_{0}=\int_{a_{+}}^{a_{-}}\phi_{h}(\phi_{0})d\phi_{0}, \label{eq:sigma0fi}
\end{equation}
where $\phi_h$ is the function defined in Eq.~(\ref{eq:primeraint0}).
This expression can be readily obtained from $\sigma_0=\int_{-\infty}^{+\infty}(\partial_{n}\phi_0)^{2}dn$ using 
$d\phi_{0}/dn=-\phi_{h}$ and $dn=-d\phi_{0}/\phi_{h}(\phi_{0})$.
The advantage of Eq.~(\ref{eq:sigma0fi}) is that it gives $\sigma_0$ directly
from the degenerate potential $V_{0}(\phi)$ without going through with profile $\phi_0(n)$.

Using the same change of variables, we can obtain all the quantities
appearing in the perturbative calculation as integrals involving the functions $\phi_{h}(\phi)$
and $\tilde{V}(\phi)$. In particular, from the first of Eqs.~(\ref{eq:Gi}) we obtain the LO values of some quantities,
\begin{equation}
	I_{0}^{(0)}=\int_{a_{+}}^{\phi_{0}}\phi_{h}(\phi)d\phi, 
	\quad
	I_{0}^{(1)}=\int_{a_{+}}^{\phi_{0}}\phi_{h}(\phi)n(\phi)d\phi,
	\quad
	\mu_{0}=\int_{a_{+}}^{a_{-}}\phi_{h}(\phi)n(\phi)^{2}d\phi.\label{eq:G0H0mu0fi}
\end{equation}
The first of these functions appears in the function $f_{1}$, Eq.~(\ref{f1}).
Using Eq.~(\ref{eq:K0}), we have
\begin{equation}
	f_{1}(\phi_{0})=\tilde{V}(\phi_{0})-\tilde{V}(a_{+})+(\Delta V_{1}/\sigma_{0})I_{0}^{(0)}(\phi_{0}).\label{eq:f1fi}
\end{equation}
Using this equation in Eq.~(\ref{eq:Ci}) and the latter in Eqs.~(\ref{eq:solfii}) and (\ref{c1}),
we obtain the quantities
\begin{equation}
	C_{1}(\phi_{0})=\int_{\phi_{*}}^{\phi_{0}}\frac{f_{1}(\phi)}{\phi_{h}(\phi)^{3}}d\phi,
	\quad 
	c_{1}=2\sigma_{0}^{-1}\int_{a_{+}}^{a_{-}} 
	C_{1}(\phi_{0})\left[V_{0}'(\phi_{0})\,n(\phi_{0})-\phi_{h}(\phi_{0})\right]d\phi_{0}.\label{eq:C1fi}
\end{equation}
and the first correction to the profile,
$ 
\phi_{1}(\phi_{0})=\phi_{h}(\phi_{0})[C_{1}(\phi_{0})+c_{1}] 
$. 
Both $\phi_{h}$ and $f_{1}$ vanish at $\phi_{0}=a_{\pm}$ (corresponding
to $n=\pm\infty$). In these limits, $\phi_{1}$ takes the values
\begin{equation}
	\phi_{1\pm}=-\frac{\tilde{V}'(a_{\pm})}{V_{0}''(a_{\pm})}=-\frac{V'(a_{\pm})}{V_{0}''(a_{\pm})},\label{eq:fi1pm}
\end{equation}
and it can be seen that we have $\partial_{n}\phi_{1}\to0$. 

From the second of Eqs.~(\ref{eq:Gi}),
we obtain the NLO integrals
\begin{equation}
	I_{1}^{(0)}(\phi_{0})=\phi_{h}(\phi_{0})\phi_{1}(\phi_{0})+\int_{a_{+}}^{\phi_{0}}\frac{f_{1}(\phi)\,d\phi}{\phi_{h}(\phi)},
	\quad
	\sigma_{1}=\int_{a_{+}}^{a_{-}}\frac{f_{1}(\phi)}{\phi_{h}(\phi)}d\phi,
	\label{eq:G1fi}
\end{equation}
where we have used Eq.~(\ref{eq:primeraintpert}), and Eq.~(\ref{mu1}) gives
\begin{equation}
	\mu_{1}=\int_{a_{+}}^{a_{-}}\frac{f_{1}(\phi)}{\phi_{h}(\phi)}n(\phi)^{2}d\phi 
	+2 \int_{a_{+}}^{a_{-}} \phi_{h}(\phi)C_{1}(\phi) n(\phi)d\phi . \label{sigmamu1}
\end{equation}
Taking into account Eqs.~(\ref{eom1}), (\ref{eom2}), and (\ref{eq:primeraintpert}), the function $\tilde{f}_{2}$, Eq.~(\ref{f2}), is given
by 
\begin{equation}
	\tilde{f}_{2} =
	\tilde{V}'(\phi_{0})\phi_{1} + {\textstyle\frac{1}{2}} V_{0}''(\phi_{0})\phi_{1}^{2} 
	- \frac{[f_{1}+V_{0}'(\phi_{0})\phi_{1}]^{2}}{2\phi_{h}^{2}}+\frac{\Delta V_{1}}{\sigma_{0}}I_{1}^{(0)}
	+\left[\frac{\Delta V_{2}}{\sigma_{0}} - \frac{\Delta V_{1} \sigma_{1}}{\sigma_{0}^2} \right] I_{0}^{(0)}  - V_{2+}.
	\label{f2tilfi}
\end{equation}
Using these equations in Eqs.~(\ref{eq:C2ab})-(\ref{eq:c2ab}), we obtain the quantities
\begin{align}
	& C_{2a}(\phi_{0}) =\int_{\phi_{*}}^{\phi_{0}}\frac{\tilde{f}_{2}(\phi)}{\phi_{h}(\phi)^{3}}d\phi,\quad C_{2b}(\phi_{0})=-\int_{\phi_{*}}^{\phi_{0}}\frac{I_{0}^{(1)}(\phi)}{\phi_{h}(\phi)^{3}}d\phi,\label{eq:C2fi}
	\\
	& c_{2a}  =\sigma_{0}^{-1} \! \int_{a_{+}}^{a_{-}} \!
	\left\{ 2\left[V_{0}'(\phi)n(\phi)-\phi_{h}(\phi)\right]C_{2a}(\phi)
	- \left[f_{1}(\phi)+V_{0}'(\phi)\phi_{1}(\phi)\right]^{2} \frac{n(\phi)}{\phi_{h}(\phi)^{3}}\right\}
	d\phi,\label{c2a}\\
	& c_{2b}  = 2\sigma_{0}^{-1}\int_{a_{+}}^{a_{-}} \left[V_{0}'(\phi)n(\phi)-\phi_{h}(\phi)\right]
	C_{2b}(\phi)d\phi \label{c2b}
\end{align}
and the second correction to the profile,
$\phi_{2}=\phi_{2a}+\phi_{2b}\partial_{n}K_{0}$,
with
\begin{equation}
	\phi_{2a}(\phi_{0})=\phi_{h}(\phi_{0})\left[C_{2a}(\phi_{0})+c_{2a}\right],
	\quad\phi_{2b}(\phi_{0})=\phi_{h}(\phi_{0})\left[C_{2b}(\phi_{0})+c_{2b}\right].\label{fi2abfi}
\end{equation}
At $\phi_{0}=a_{\pm}$ (i.e., $n=\pm\infty$), $\phi_{2}$ takes the
values
\begin{equation}
	\phi_{2\pm}=-\frac{\tilde{V}''(a_{\pm})\phi_{1\pm}+\frac{1}{2}V_{0}'''(a_{\pm})\phi_{1\pm}^{2}}{V_{0}''(a_{\pm})}.
\end{equation}
We can also write Eqs.~(\ref{eq:sigma2sep}) and (\ref{eq:sigmamu2}) as
\begin{align}
	\tilde{\sigma}_{2} & =\int_{a_{+}}^{a_{-}} \left[\frac{\left[f_{1}(\phi)+V_{0}'(\phi)\phi_{1}(\phi)\right]^{2}}{\phi_{h}^{2}(\phi)}
	+\tilde{f}_{2}(\phi)\right] \frac{d\phi}{\phi_{h}(\phi)}. \label{sigma2tilfi} 
	\\
	\tilde{\mu}_{2} & = \int_{a_{+}}^{a_{-}} \left\{ \left[\frac{\left[f_{1}(\phi) + V_{0}'(\phi)\phi_{1}(\phi)\right]^{2}}{\phi_{h}^{3}(\phi)} + \frac{\tilde{f}_{2}(\phi)}{\phi_{h}(\phi)}\right]n(\phi)^{2}
	+ 2\phi_h(\phi)C_{2a}(\phi) \,n(\phi) \right\} d\phi, \label{mu2tilfi}
	\\
	\alpha & =\int_{a_{+}}^{a_{-}} \phi_{h}(\phi) \left[2C_{2b}(\phi)\,n(\phi)- {\textstyle\frac{1}{3}}n(\phi)^{4} \right]d\phi. \label{alfafi}
\end{align}

Let us now consider the potential $V_0$ of Eq.~(\ref{eq:V0}) and the linear $\tilde{V}=c\phi$.
We begin by calculating the LO field profile $\phi_0(n)$ from Eqs.~(\ref{eq:primeraint0})-(\ref{eq:nast}).
The result does not depend on the integration constant $\phi_{*}$, and we shall take $\phi_{*}=(a_{+}+a_{-})/2$
to take advantage of the symmetry of $V_0$. 
We also define the parameter $a=(a_{-}-a_{+})/2$ and
we  make the changes of variables $x=(\phi-\phi_{*})/a$,
$x_{0}=(\phi_{0}-\phi_{*})/a$ in the integrals in Eqs.~(\ref{eq:fi0})-(\ref{eq:nast}). 
Since the
integration range of $\phi$ and $\phi_{0}$ is the interval $[a_{+},a_{-}]$, 
the variables $x$ and $x_{0}$ vary between $-1$ and $+1$.
Thus, the function $\phi_h$  is given by
$\phi_{h}=\sqrt{\lambda/2}a^{2}(1-x_{0}^{2}) $. 
After this change of variable in Eq.~(\ref{eq:fi0}), it is apparent
by symmetry that $n_{*}=0$, and we obtain
\begin{equation}
	n=-\frac{a^{-1}}{\sqrt{2\lambda}}\log\left(\frac{1-x_{0}}{1+x_{0}}\right).\label{eq:npot}
\end{equation}
Inverting this expression, we obtain $x_{0}=-\tanh(\sqrt{\lambda/2}\,a\,n)$.

The LO quantities (\ref{eq:sigma0fi})-(\ref{eq:G0H0mu0fi}) give
$\sigma_{0}=\frac{4}{3}\sqrt{\lambda/2}\,a^{3}$,  $\mu_{0}=\frac{\pi^{2}-6}{9}\sqrt{2/\lambda}\,a$,
\begin{align}
	I_{0}^{0}(x_{0}) & =a^{3}\sqrt{\lambda/2}\left(2/3+x_{0}-x_{0}^{3}/3\right),\\
	I_{0}^{1}(x_{0}) & =
	\frac{1}{6}a^{2}\left[x_{0}^{2}-1+x_{0}(x_{0}^{2}-3)\log\left(\frac{1+x_{0}}{1-x_{0}}\right)
	-2\log\left(\frac{1-x_0^{2}}{4}\right)\right]. \label{H0pot}
\end{align}
For $\tilde{V}=c\phi$, Eq.~(\ref{V1})
gives $\Delta V_{1}=-2ca$, and we obtain
$f_{1}=-(ca/2)x_{0}(1-x_{0}^{2})$ and $\phi_{1}=-c/(2a^{2}\lambda)$.
In this case, we have 
$V_{0}''(a_{+})=V_{0}''(a_{-})$ 
and  the derivative $\tilde{V}'$ is a constant.
As a consequence, Eq.~(\ref{V1}) gives $\Delta V_{2}=0$,
and we obtain
\begin{equation}
	\tilde{f}_{2}(x_{0})=\frac{3c^{2}}{8\lambda a^{2}}(x_{0}^{2}-1),
	\quad
	C_{2a}(x_{0}) =\frac{3\sqrt{2}c^{2}}{16\lambda^{5/2}a^{7}}
	\left[\frac{2x_{0}}{x_{0}^{2}-1}-\log\left(\frac{1+x_{0}}{1-x_{0}}\right)\right] ,
\end{equation}
and
\begin{multline}
	C_{2b}(x_{0})  = -\frac{\sqrt{2}a^{-3}}{48\lambda^{3/2}} 
	\bigg\{ \frac{4x_{0}}{(1-x_{0}^{2})^{2}} \left[ (1-6\log2)x_{0}^{2} + 10\log2 -1 \right]  
	\\
	-2\log(1-x_{0}) \left[\frac{x_0(5x_{0}+4)-3}{(1+x_0)^{2}}
	+3\log2\right]+2\log(1+x_0)
	\left[\frac{x_0(5x_0-4)-3}{(1-x_0)^{2}}+3\log2\right] 
	\\
	+3\log^{2}(1-x_{0})-3\log^{2}(1+x_{0}) +6\textrm{Li}_{2}
	\left(\frac{1+x_{0}}{2}\right)-6\textrm{Li}_{2}\left(\frac{1-x_{0}}{2}\right)\bigg\} ,
\end{multline}
where $\textrm{Li}_{2}(z)=\sum_{k=1}^{\infty}z^{k}/k^{2}=-\int_{0}^{z}\log(1-t)dt/t$
is the dilogarithm function \cite{abramowitz}. 
The integrals (\ref{c2a})-(\ref{c2b}) give $c_{2a}=c_{2b}=0$ since their integrands are odd after the change of variables.
The first term in the integrands of (\ref{sigma2tilfi})-(\ref{mu2tilfi})
vanishes in this case, which is more easily seen in the expressions
(\ref{eq:Gi}), and we obtain
\begin{equation}
	\tilde{\sigma}_{2}=-\frac{3\sqrt{2}}{4\lambda^{3/2}}\frac{c^{2}}{a^{3}},\quad\tilde{\mu}_{2}=-\frac{\sqrt{2}\left(\pi^{2}-6\right)}{24\lambda^{5/2}}\frac{c^{2}}{a^{5}},
	\quad\alpha=0.346305\frac{(2/\lambda)^{3/2}}{3a}.
\end{equation}
Thus, for instance, Eq.~(\ref{eq:sigmasep}) gives
\begin{equation}
	\frac{\sigma}{\sigma_{0}} = 1-\frac{9c^{2}}{8\lambda^{2}a^{6}}
	-\left(\frac{\pi^{2}}{6}-1\right)\frac{\partial_{n}K_{0}|_{0}}{\lambda a^{2}} . \label{eq:sigmatotpot}
\end{equation}

\bibliographystyle{jhep}
\bibliography{papers}
\end{document}